\documentclass[12pt]{iopart}

\usepackage{tabularx}
\expandafter\let\csname equation*\endcsname\relax
\expandafter\let\csname endequation*\endcsname\relax
\usepackage{amsmath}
\usepackage{amssymb}
\usepackage{graphicx}
\usepackage{dcolumn}
\usepackage{ulem}
\usepackage{bm}

\def\bea{\begin{eqnarray}}
\def\eea{\end{eqnarray}}

\def\be{\begin{eqnarray}}
\def\ee{\end{eqnarray}}

\newcommand{\onlinecite}[1]{\hspace{-1 ex} \nocite{#1}\cite{#1}}

\begin{document}

\title{Fulde-Ferrell-Larkin-Ovchinnikov phase and Symmetry of Fermi Surface in Two-dimensional Spin-Orbit Coupled Fermi Gas}

\author{Zhen Zheng}
\address{Key Laboratory of Quantum Information, University of Science and
Technology of China, Hefei, Anhui, 230026, People's Republic of China}

\author{Ming Gong}
\address{Department of Physics and Centre for Quantum Coherence, The Chinese University of Hong Kong, Shatin, N.T., Hong Kong, China}
\ead{skylark.gong@gmail.com}

\author{Yichao Zhang}
\address{Key Laboratory of Quantum Information, University of Science and
Technology of China, Hefei, Anhui, 230026, People's Republic of China}

\author{Xubo Zou}
\address{Key Laboratory of Quantum Information, University of Science and
Technology of China, Hefei, Anhui, 230026, People's Republic of China}
\ead{xbz@ustc.edu.cn}

\author{Chuanwei Zhang}
\address{Department of Physics, The University of Texas at Dallas, Richardson,
TX, 75080 USA}
\ead{chuanwei.zhang@utdallas.edu}

\author{Guangcan Guo}
\address{Key Laboratory of Quantum Information, University of Science and
Technology of China, Hefei, Anhui, 230026, People's Republic of China}

\begin{abstract}
We show that the combination of spin-orbit coupling and in-plane Zeeman
field in a two-dimensional degenerate Fermi gas can lead to a larger
parameter region for Fulde-Ferrell-Larkin-Ovchinnikov (FFLO) phases than
that using spin-imbalanced Fermi gases. The resulting FFLO superfluids are
also more stable due to the enhanced energy difference between FFLO and
conventional Bardeen-Cooper-Schrieffer (BCS) excited states. We clarify
the crucial role of the symmetry of Fermi surface on the formation of finite
momentum pairing. The phase diagram for FFLO
superfluids is obtained in the BCS-BEC crossover region and possible
experimental observations of FFLO phases are discussed.
\end{abstract}

\pacs{05.30.Fk, 74.20.Fg, 74.25.Dw, 67.85.-d}
\maketitle

% using the pacs number in 2012
% 74.20.Fg: BCS theory and its development
% 71.10.Ca: Electron gas, Fermi gas
% 05.30.Fk: Fermion systems and electron gas
% 74.70.Tx: Heavy-fermion superconductors
% 74.25.Dw: Superconductivity phase diagrams
% 37.10.Gh: Atom traps and guides
% 71.10.Fd: Lattice fermion models (Hubbard model, etc.)
% 67.85.-d: Ultracold gas, trapped gas
% 71.70.Ej: Spin-orbit coupling, Zeeman and Stark splitting, Jahn-Teller effect

\section{Introduction}

In 1964, just shortly after the great success of Bardeen-Cooper-Schrieffer
(BCS) theory for superconductivity \cite{BCS}, Fulde and Ferrell (FF) \cite%
{Ferrell64}, and Larkin and Ovchinnikov (LO) \cite{Larkin64,Larkin65}
independently demonstrated that a new type of superconducting state, which
is characterized by Cooper pairs with nonzero total momentum, may exist in certain 
regime of a clean superconductor under a strong magnetic field. The order parameters in real space
for these two superconductor read as
\begin{equation}
\Delta_{\text{FF}}({\bf x}) = \Delta e^{i{\bf Q}\cdot {\bf x}}, \quad
\Delta_{\text{LO}}({\bf x}) = \Delta \cos({\bf Q}\cdot {\bf x}).
\end{equation}
The superconducting state is now known as the FFLO superconductor or inhomogeneous superconductor.
For conventional BCS superconductors\cite{BCS},
the pairing takes place between electrons with opposite momentum and opposite spin,
i.e., $\mathbf{k}\uparrow $ and $-\mathbf{k}\downarrow $. Therefore when the
magnetic field exceeds certain critical value, the superconductivity is
destroyed due to Pauli paramagnetic depairing effect. As a consequence,
magnetism and superconductivity generally cannot coexist for the BCS type-I superconductor.
The physics is totally different for FFLO phases because these
two different orders naturally coexist; more precisely, the FFLO phase
arises from the interplay between magnetism and superconductivity. This
important feature makes the FFLO phase a central concept for understanding
many exotic phenomena in different physics branches, ranging from
unconventional solid state superconductors (e.g., layered \cite{Buzdin,
Croitoru}, heavy-fermion \cite{Yuji,Gloos,Bianchi}, organic \cite{Singleton,
Lortz} superconductors, \textit{etc}.), to chiral quark matter in quantum chromodynamics (QCD),
and to neutron star glitches in astrophysics \cite{Casalbuoni,Alford}. In the past several decades, great efforts have been
made to unveil this novel quantum phase, and a lot of exotic signatures that may be related to the FFLO phase have been observed,
for instance, the anisotropic thermal conductivity \cite{Capan}, specific heat\cite{Bianchi}, nuclear magnetic resonance\cite{NMR1, NMR2, NMR3},
and ultrasound velocities\cite{Ultrasound} have been ascribed to the formation of FFLO superconductor in the heavy fermion superconductor CeCoI$_5$.
However, until now, clear, unambiguous and direct experimental evidences for the existence of
FFLO phases are still lacking \cite{Casalbuoni,Houzet}. There are several reasons for that:
the existence of FFLO phase requires very stringent conditions; the direct
probing of periodic oscillation of the order parameter is challenging; and
the disorder effects in the superconductor induce strong scattering between
different momenta that destroys the superconducting pairing \cite{Alsam,
JXZhu}.

The recent experimental advances of population-imbalanced ultracold Fermi
gases may have the potential to elucidate this long-sought problem. The
ultracold atomic system possesses some remarkable advantages over their
counterpart in solid state systems due to its high controllability and
tunability \cite{Feshbach1, Feshbach2, Feshbach3}. The experimental
parameters in ultracold atoms can be tuned in realistic experiments.
Furthermore, the system can be made disorder free, which if
necessary, can be introduced to the system in a controllable manner \cite%
{disorder1, disorder2, disorder3}. On the experimental side, the
superfluidity of the Fermi gas can be characterized by the generation of
vortices when the gas is rotated\cite{Zwierlein05}, and the momentum of the
Cooper pair in the FFLO phase can be directly probed using the
time-of-flight imaging \cite{Sheehy, TOF2}, while in solid state the direct
observation of the FFLO phase and its Cooper pair momentum is challenging.
Unfortunately, this system still have two major obstacles hinder the observation
of FFLO phase in recently experiments. Firstly, the FFLO phase only exists in an
extremely narrow parameter regime in 2D and 2D (see Fig. \ref{fig-phase}a) degenerate Fermi
gases \cite{Sheehy, Hu}, therefore in experiments the FFLO phase is generally missed out.
For instance, in recent experiments with population-imbalanced Fermi gases \cite%
{Zwierlein06a, Zwierlein06b} only the phase transition from BCS superfluids
to normal gases has been observed. While in another experiment \cite%
{Hulet1} the phase separation phase, which is also known as the breached
pair \cite{Vicent}, has been observed. Secondly, the energy difference between
FFLO ground state and BCS excited state is much smaller than the
temperature, therefore even the parameters for the FFLO state have been reached,
the Fermi gas is still too hot to reach the ground state.

The above two obstacles can be overcome using spin-orbit (SO) coupled degenerate Fermi gases with an in-plane Zeeman
field. Here the SO coupling is a central ingredient in modern physics, because it is essential to a number of
important concepts in condensed matter physics, ranging from spin Hall effect \cite{SHE1, SHE2}, anomalous
quantum Hall effect \cite{AHE1, AHE2}, and topological insulators \cite{TI1, TI2, TI3}.
In solid materials, the SO coupling is induced by inversion symmetry of bulk or structure \cite{Winkle}.
However, in cold atom systems, the SO coupling is induced by Raman coupling between
hyperfine states \cite{Ruseckas,Xiongjun,xiongjun2,CWZhang,Campbell,Jay,non-Abelian}, therefore in principle, different types
of SO coupling can be created by carefully choosing different laser configurations. Experimentally, the one
dimensional SO coupling has been realized using Raman coupling between
hyperfine states for both Bose and Fermi gases \cite{Spielmana,Spielmanb,Pan,Qu,Jing,Zwierlein12}, while the in-plane Zeeman
field naturally exists in this system. Here we show that the combination of
a Rashba-type of SO coupling and an in-plane Zeeman field can support FFLO
superfluids with a unique FFLO vector in a 2D degenerate Fermi gas. The
required Zeeman field or the population imbalance can be extremely small
with realistic experimental parameters. The driving mechanism for the FFLO
superfluids is the interplay between deformation of Fermi surface and
superconducting order \cite{Zhen}, thus should be in stark constrast to the
physics in original FFLO superconductor \cite{Ferrell64, Larkin64,Larkin65}. Recently, there
are already several related works showing that the FFLO superfluids can be observed
with different SO couling and Zeeman fields \cite{SOFF1, SOFF2, SOFF3, SOFF4, SOFF5, SOFF6},
and all of them belong to the scope of this new driven mechanism. Here in this work, we provide
a comprehensive understanding for the formation of FFLO superfluids in the SO coupled Fermi gas
from the symmetry of Fermi surface.

The rest of this work is organized as following. We present our mean field
treatment of the SO coupled degenerate Fermi gas with an in-plane Zeeman
field in Sec. \ref{sec-measurementodel}, and we discuss the basic particle-hole
symmetry of the effective Hamiltonian in Sec. \ref{sec-phsymmetry}. The rotational
symmetry breaking due to the in-plane Zeeman field is presented in Sec. \ref%
{sec-FSsymmetry}. The numerical details for the FFLO superfluid are presented
in Sec. \ref{sec-numericaldetails}. We plot the phase diagram and discuss its basic
properties in Sec. \ref{sec-phasediagram} and we plot the free energy landscape in
Sec. \ref{sec-Flandscape}. We discuss the measurement of the FFLO
phase in Sec. \ref{sec-measurement}. The notable differenes between cold atom systems
and solid materials are discussed in Sec. \ref{sec-vssolids}.
At last we conclude in Sec. \ref{sec-conclusion}.

\section{Physical Model}
\label{sec-measurementodel}

We consider a 2D degenerate Fermi gas with Rashba-type SO coupling and an
in-plane Zeeman field. The 2D degenerate Fermi gases can be constructed by
applying a strong standing wave along the third direction, and have been
realized in recent experiments \cite{Martiyanov}. The 2D SO coupled Fermi
gases can be described as\cite{CWZhang2, Gong11}
\begin{equation}
H=\sum_{\mathbf{k}\sigma \sigma ^{\prime }}c_{\mathbf{k},\sigma }^{\dagger
}[\xi _{\mathbf{k}\sigma }+\alpha (k_{x}\sigma _{y}-k_{y}\sigma
_{x})-h\sigma _{x}]c_{\mathbf{k},\sigma ^{\prime }}+V_{\text{int}},
\end{equation}%
where $\alpha $ is the SO coupling strength, $\sigma _{x}$ and $\sigma _{y}$ are the Pauli operators,
$\xi _{\mathbf{k}\sigma }={\frac{k^{2}}{2m}}-\mu $, and $\mathbf{k}=(k_{x},k_{y})$. The one dimensional
SO coupling has been realized in Fermionic cold atoms\cite{Jing,Zwierlein12}. The schemes to the realization
of two dimensional SO coupling are similar, but requires more complicated laser beams, see recent
works\cite{CWZhang,Campbell,Jay}.

The last term corresponds to the $s$-wave scattering interaction
\begin{equation}
V_{\text{int}}=g\sum_{\mathbf{p}_{1}+\mathbf{p}_{2}=\mathbf{p}_{3}+\mathbf{p}%
_{4}}c_{\mathbf{p}_{1},\uparrow }^{\dagger }c_{\mathbf{p}_{2},\downarrow
}^{\dagger }c_{\mathbf{p}_{3},\downarrow }c_{\mathbf{p}_{4},\uparrow },
\label{eq-Vint}
\end{equation}%
where $\mathbf{p}_{1}+\mathbf{p}_{2}=\mathbf{p}_{3}+\mathbf{p}_{4}$ due to
the conservation of the total momentum during the scatting process, $g$ is
the scattering interaction strength.

When atoms form Cooper pairs with a finite total momentum $\mathbf{Q}$, the
scattering process in Eq. \ref{eq-Vint} can be simplified with $\mathbf{p}%
_{1}=\mathbf{k}+\mathbf{Q}/2$, $\mathbf{p}_{2}=-\mathbf{k}+\mathbf{Q}/2$, $%
\mathbf{p}_{3}=\mathbf{p}+\mathbf{Q}/2$, and $\mathbf{p}_{4}=-\mathbf{p}+%
\mathbf{Q}/2$. When $\mathbf{Q}=0$, the Cooper pairs are formed between two
atoms with opposite momentum and opposite spin, and we recover the conventional BCS
type pairing. Denote $\beta _{\mathbf{p}}=gc_{\mathbf{p}+\mathbf{Q}%
/2,\downarrow }c_{-\mathbf{p}+\mathbf{Q}/2,\uparrow }$, the interaction term
can be written as
\begin{equation}
V_{\text{int}}=g\sum_{\mathbf{k},\mathbf{p}}c_{\mathbf{k%
}+\mathbf{Q}/2,\uparrow }^{\dagger }c_{-\mathbf{k}+\mathbf{Q}/2,\downarrow
}^{\dagger }c_{\mathbf{p}+\mathbf{Q}/2,\downarrow }c_{-\mathbf{p}+\mathbf{Q}%
/2,\uparrow }=\sum_{\mathbf{k},\mathbf{p}}{\frac{\beta _{\mathbf{k}%
}^{\dagger }\beta _{\mathbf{p}}}{g}}.
\end{equation}
This interaction term can be decoupled using the standard mean-field method
\begin{equation}
\beta _{\mathbf{k}}^{\dagger }\beta _{\mathbf{p}}\rightarrow \langle \beta _{%
\mathbf{k}}^{\dagger }\rangle \beta _{\mathbf{p}}+\beta _{\mathbf{k}%
}^{\dagger }\langle \beta _{\mathbf{p}}\rangle -\langle \beta _{\mathbf{k}%
}^{\dagger }\rangle \langle \beta _{\mathbf{p}}\rangle ,
\label{eq-decoupling}
\end{equation}%
where the order parameter in the momentum space $\Delta =$ $\sum_{\mathbf{p}%
}g\langle c_{\mathbf{p}+\mathbf{Q}/2,\downarrow }c_{-\mathbf{p}+\mathbf{Q}%
/2,\uparrow }\rangle $. The interaction term now reduces to
\begin{equation}
V_{\text{int}}=\sum_{\mathbf{k}}\Delta \beta _{\mathbf{k}}^{\dagger }+\Delta
^{\ast }\beta _{\mathbf{k}}-{\frac{|\Delta |^{2}}{g}.}
\end{equation}%
Notice that $c_{\mathbf{p}\sigma }=\int d\mathbf{x}c_{\sigma }(\mathbf{x}%
)e^{-i\mathbf{p}\cdot \mathbf{x}}$, thus we have the paring in the real
space,
\begin{eqnarray}
\langle c_{\uparrow }(\mathbf{x})c_{\downarrow }(\mathbf{y})\rangle &=&\int d%
\mathbf{p}d\mathbf{p}^{\prime }\langle gc_{\mathbf{p},\uparrow }c_{\mathbf{p}%
^{\prime },\downarrow }\rangle e^{i(\mathbf{p}\cdot \mathbf{x}+\mathbf{p}%
^{\prime }\cdot \mathbf{y})}  \notag \\
&=&\Delta \delta (\mathbf{x}-\mathbf{y})e^{i\mathbf{Q}\cdot \mathbf{x}}.
\end{eqnarray}%
We see the finite momentum pairing in the momentum space corresponds to a
spatially modulated pairing in the real space. The $\delta $-function arises from
the contact interaction. The above pairing breaks the time-reversal
symmetry, which is in consistent with our model because a Zeeman field is
applied.

Using the standard Bogoliubov transformation, the Hamiltonian can be
written as%
\begin{equation}
H={\frac{1}{2}}\sum_{\mathbf{k}}\psi _{\mathbf{k},\mathbf{Q}}^{\dagger }H_{%
\text{eff}}\psi _{\mathbf{k},\mathbf{Q}}-{\frac{|\Delta |^{2}}{g}}+{\frac{1}{%
2}}\sum_{\mathbf{k},\sigma }\xi _{\mathbf{k}\sigma },  \label{eq-psikQ}
\end{equation}%
where the effective Hamiltonian reads as,
\begin{equation}
H_{\text{eff}}=%
\begin{pmatrix}
\mathcal{K}(\mathbf{k}) & \Delta I_{2\times 2} \\
\Delta ^{\dagger }I_{2\times 2} & -\sigma _{y}\mathcal{K}^{\ast }(-\mathbf{k}%
)\sigma _{y}%
\end{pmatrix}
\label{eq-Heff}
\end{equation}%
with
\begin{equation}
\mathcal{K}(\mathbf{k})=%
\begin{pmatrix}
\xi _{\mathbf{k}+\mathbf{Q}/2,\uparrow } & h-\alpha R(\mathbf{k}) \\
h-\alpha R^{\ast }(\mathbf{k}) & \xi _{\mathbf{k}+\mathbf{Q}/2,\downarrow }%
\end{pmatrix}%
,
\end{equation}%
$R(\mathbf{k})=(\mathbf{k}+\mathbf{Q}/2)_{x}+i(\mathbf{k}+\mathbf{Q}/2)_{y}$%
, and $I_{2\times 2}=\text{diag}(1,1)$. The basis defined in Eq. \ref%
{eq-psikQ} is $\psi _{\mathbf{k},\mathbf{Q}}=(c_{\mathbf{k}+\mathbf{Q}%
/2,\uparrow },c_{\mathbf{k}+\mathbf{Q}/2,\downarrow },c_{-\mathbf{k}+\mathbf{%
Q}/2,\downarrow }^{\dagger },-c_{-\mathbf{k}+\mathbf{Q}/2,\uparrow
}^{\dagger })^{T}$. The minus sign in the last term of the basis is used to
achieve the $\Delta I_{2\times 2}$ type off-diagonal term in Eq. \ref%
{eq-Heff}.

The thermodynamical potential at zero temperature reads as
\begin{equation}
\Omega =-{\frac{\Delta ^{2}}{g}}+{\frac{1}{2}}\sum_{\mathbf{k}\sigma }\xi _{%
\mathbf{k}\sigma }+{\frac{1}{2}}\sum_{\mathbf{k},\lambda }E_{\lambda }\Theta
(-E_{\lambda }),  \label{eq-omega}
\end{equation}%
where the Heaviside step function
\begin{equation}
\Theta (x)=%
\begin{cases}
1, & \quad x\geq 0 \\
0, & \quad x<0%
\end{cases}%
.
\end{equation}%
$E_{\lambda }$, $\lambda =$1, 2, 3 and 4, are the eigenvalues of the
effective Hamiltonian $H_{\text{eff}}$, whose exact expressions are too
complex to be presented here. We therefore see the standard mean-field decoupling used in
Eq. \ref{eq-decoupling} actually corresponds to the Hubbard-Stratanovich
transformation in the quantum field theory. We use the mean field theory as
the main theoretical tool in this work because it provides a more
transparent description of the FFLO physics.

The effective scattering interaction $g$ in Eq. \ref{eq-omega} in a 2D Fermi
gas should be regularized through \cite{Gang}
\begin{equation}
{\frac{1}{g}}=-\sum_{\mathbf{k}}{\frac{1}{\mathbf{k}^{2}/m+E_{b}}},
\label{eq-reg}
\end{equation}%
where the binding energy $E_{b}$ can be tuned by varying the $s$-wave
scattering length through Feshbach resonance \cite{Feshbach1, Feshbach2,
Feshbach3}.

\section{Particle-Hole Symmetry}

\label{sec-phsymmetry}

\begin{figure}[tbp]
\centering
\includegraphics[width=5in]{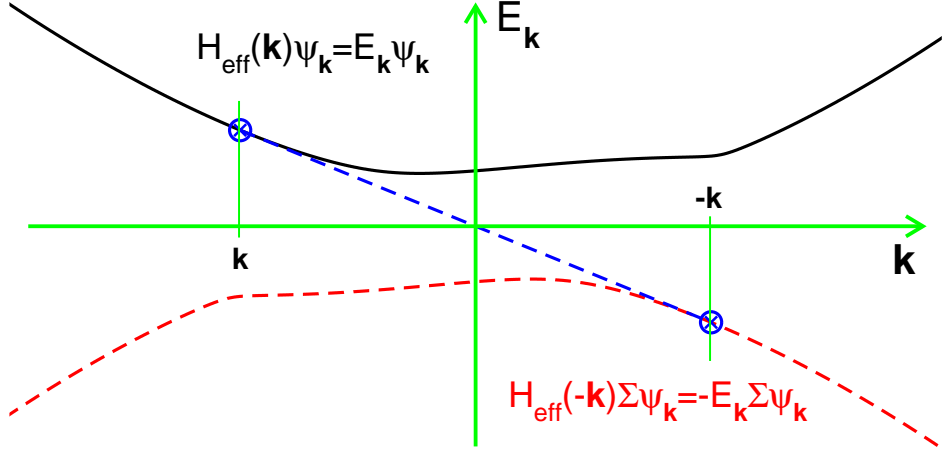}
\caption{Particle-hole symmetry in our model. The particle-hole symmetry
establishes the direct relation between $\protect\psi _{\mathbf{k}}$ for $H_{%
\text{eff}}(\mathbf{k})$ with eigenvalues $E_{\mathbf{k}}$ to $\Sigma
\protect\psi _{\mathbf{k}}$ for $H_{\text{eff}}(-\mathbf{k})$ with
eigenvalues $-E_{\mathbf{k}}$. Due to the lack of chiral symmetry%
\protect\cite{Schnyder}, the eigenvalues of $H_{\text{eff}}(\mathbf{k})$ may
not appear in pairs. The particle-hole correspondence is shown via dashed
line.}
\label{fig-ph}
\end{figure}

The particle-hole operator for the SO coupled Fermi gases can be written as $%
\Sigma =\Lambda K$, where $\Lambda =\sigma _{y}\tau _{y}$ ($\sigma _{y}$ is
the Pauli spin matrix and $\tau _{y}$ is the Nambu particle-hole matrix),
and $K$ represents the complex conjugate operator. Obviously, $\Lambda $ is a unitary operator, and
$\Lambda =\Lambda ^{-1}$.  It is easy to check $\Sigma ^{2}=\Lambda K\Lambda K=\Lambda \Lambda ^{\ast
}=\Lambda ^{2}=1$. Moreover, the formation of FFLO phase breaks the time-reversal
symmetry, therefore the system belongs to the symmetry class $D$, see
Ref. \onlinecite{Schnyder}. The particle-hole operator has the following basic property,
\begin{eqnarray}
\Sigma H_{\text{eff}}(\mathbf{k})\Sigma ^{-1} &=&\Lambda H_{\text{eff}%
}^{\ast }(\mathbf{k})\Lambda  \notag \\
&=&\Lambda
\begin{pmatrix}
\mathcal{K}^{\ast }(\mathbf{k}) & \Delta ^{\dagger }I_{2\times 2} \\
\Delta I_{2\times 2} & -\sigma _{y}\mathcal{K}(-\mathbf{k})\sigma _{y}%
\end{pmatrix}%
\begin{pmatrix}
0 & -i\sigma _{y} \\
i\sigma _{y} & 0%
\end{pmatrix}
\notag \\
&=&%
\begin{pmatrix}
0 & -i\sigma _{y} \\
i\sigma _{y} & 0%
\end{pmatrix}%
\begin{pmatrix}
i\Delta ^{\dagger }\sigma _{y} & -i\mathcal{K}^{\ast }(\mathbf{k})\sigma
_{y} \\
-i\sigma _{y}\mathcal{K}(-\mathbf{k}) & -i\Delta \sigma _{y}%
\end{pmatrix}
\notag \\
&=&%
\begin{pmatrix}
-\mathcal{K}(-\mathbf{k}) & -\Delta I_{2\times 2} \\
-\Delta ^{\dagger }I_{2\times 2} & \sigma _{y}\mathcal{K}^{\ast }(\mathbf{k}%
)\sigma _{y}%
\end{pmatrix}
\notag \\
&=&-H_{\text{eff}}(-\mathbf{k}).
\end{eqnarray}%
Here $\Sigma $ establishes a one-to-one correspondence between $\mathbf{k}$
and $\mathbf{-k}$, therefore if $\psi _{\mathbf{k}}=(u(\mathbf{k}),v(\mathbf{%
k}))^{T}$ is an eigenvector of the Hamiltonian $H_{\text{eff}}(\mathbf{k})$
with energy $E(\mathbf{k})$, then $\psi _{\mathbf{k}}^{\prime }=\Sigma \psi
_{\mathbf{k}}=(\sigma _{y}v^{\ast }(\mathbf{k}),-\sigma _{y}u^{\ast }(%
\mathbf{k}))^{T}$ is the eigenvector of the Hamiltonian $H_{\text{eff}}(-%
\mathbf{k})$ with energy $-E(\mathbf{k})$ (see in Fig. \ref{fig-ph}). The particle-hole symmetry does not automatically ensure that
the eigenvalues appear in pairs ($E$, $-E$) because the system lacks the
chiral symmetry \cite{Schnyder}. This is a direct consequence of the
inversion symmetry breaking (see Sec. \ref{sec-FSsymmetry}) of our model in
the presence of an in-plane Zeeman field. The trace of the effective
Hamiltonian gives,
\begin{equation}
\text{tr}(H_{\text{eff}}(\mathbf{k}))=\sum_{\lambda }E_{\lambda }={\frac{2%
\mathbf{k}\cdot \mathbf{Q}}{m}},
\label{eq-elambda}
\end{equation}%
therefore for the FFLO phase, the eigenvalues never appear with pairs. Eq. \ref{eq-elambda}
is essential for us to understand the band structures of the FFLO superfluid. With only in-plane
Zeeman field, only trivial phase can be observed, see discussion in Ref. \onlinecite{NC}.

\section{Symmetry of Fermi Surface}

\label{sec-FSsymmetry}

The symmetry of Fermi surface is essential to understand the properties of different
quantum phases and their signature in the time-of-flight imaging, which is the basic
motivation of this work. The Rashba type SO coupling, $V_{\text{so}}=\alpha
(k_{x}\sigma _{y}-k_{y}\sigma _{x})$\textbf, is
invariant under the simultaneous rotation of the momentum and spin in the $xy$ plane,
\begin{equation}
\begin{pmatrix}
-k_{y}^{\prime } \\
k_{x}^{\prime }%
\end{pmatrix}%
=U%
\begin{pmatrix}
-k_{y} \\
k_{x}%
\end{pmatrix}%
,\quad
\begin{pmatrix}
\sigma _{x}^{\prime } \\
\sigma _{y}^{\prime }%
\end{pmatrix}%
=U%
\begin{pmatrix}
\sigma _{x} \\
\sigma _{y}%
\end{pmatrix}%
,
\end{equation}%
where $U$ is the SO(2) rotation matrix,
\begin{equation}
U=%
\begin{pmatrix}
\cos (\theta ) & \sin (\theta ) \\
-\sin (\theta ) & \cos (\theta )%
\end{pmatrix}%
.  \label{eq-SO2}
\end{equation}%
The SO(2) rotation matrix does not change the magnitude of the momentum, thus
$\xi _{\mathbf{k}%
^{\prime }\sigma }=\xi _{\mathbf{k}\sigma }$
 is also invariant under this transformation. Meanwhile, by defining
$\sigma _{z}^{\prime }=\sigma_{z}$,
the new Pauli matrices $\sigma _{x,y,z}^{\prime }$ satisfy the standard commutation relation
\begin{equation}
\lbrack \sigma _{a}^{\prime },\sigma _{b}^{\prime }]=2i\sum_{c}\varepsilon
_{abc}\,\sigma _{c}^{\prime },\quad \{\sigma _{a}^{\prime },\sigma
_{b}^{\prime }\}=2\delta _{ab},
\end{equation}%
with $\varepsilon_{abc}$ the Levi-Civita symbol and $\delta _{ab}$ the Kronecker delta.

The SO(2) symmetry may breakdown in the present of both Rashba and Dresselhaus SO coupling. However, in
this case, the Fermi surface still has inversion symmetry, which means that the eigenvalues of single particle
Hamiltonian have the basic property $E_{{\bf k}\sigma} = E_{-{\bf k}\sigma}$ for any ${\bf k}$ and $\sigma$.
This symmetry is unbroken by out-of-plane Zeeman field. The inversion symmetry of Fermi surface is most
relevant to the physics in this work, and
it is exact this symmetry ensures that the BCS phase instead of FFLO phase is more energetically favorable
in the present of out-of-plane Zeeman field. An intuitive understanding of this result is that for
any state with momentum ${\bf k}$, we can always find another degenerate state with opposite momentum
at the same band, thus we have BCS phase. The SO coupling here plays the role of inducing pairing at the
same band.

The inversion symmetry is broken by in-plane Zeeman field because the rotational in Eq. \ref{eq-SO2}
results in the following transformation,
\begin{equation}
\sigma _{x}\rightarrow \cos (\theta )\sigma _{x}+\sin (\theta )\sigma _{y}.
\end{equation}
Physically, it means that it is impossible to find two degenerate states with opposite momentum at the same band.
This anisotropy effect also lead to a unique FFLO vector $\mathbf{Q}$ for the FFLO superfluid, which
is one of the key point of our proposal in Ref. \onlinecite{Zhen}. The unique $\mathbf{Q}$
makes the detection of the FFLO vector much easier in realistic experiments, see more discussions in
Sec. \ref{sec-measurement}. This picture is quite general and for this basic reason, the  FFLO phase in this work
can also be realized using other types of SO coupling\cite{SOFF1,SOFF2,SOFF3,SOFF4,SOFF5, SOFF6}.

\begin{figure}[t]
\centering
\includegraphics[width=5.5in]{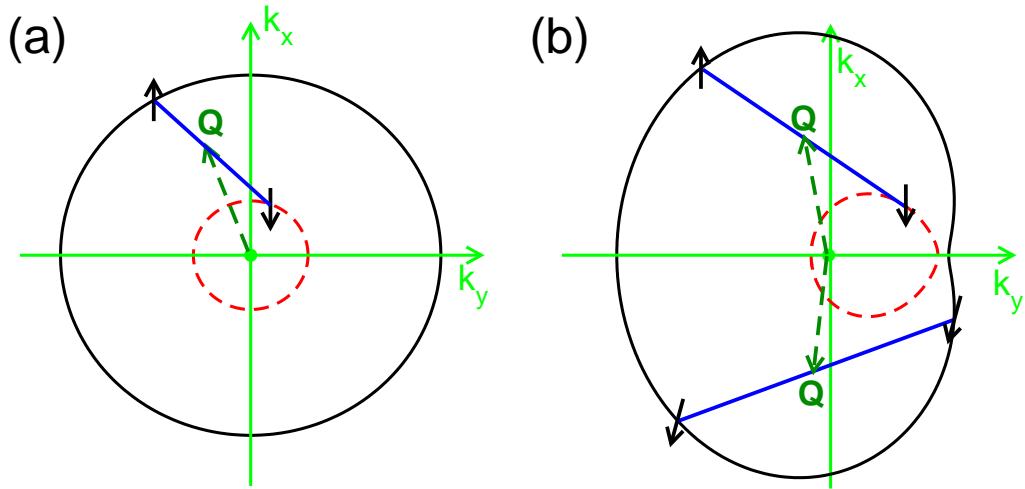}
\caption{(Color online) Basic physical picture for the emergence of FFLO
phases in ultracold atomic systems. (a) shows the creation of FFLO phases with
a Zeeman field, which is equivalent to the physics by controlling the
population imbalance. The mismatched Fermi surface makes the pairing with
opposite momentum and spin challenging, thus in some parameter regime the
FFLO phase becomes energetically favorable. This mechanism is inefficient
because the Pauli paramagnetic depairing effect under strong Zeeman field
destroys the Cooper pairs, thus the FFLO phase only survives in a very
narrow parameter regime. (b) shows a new mechanism for the generation of
FFLO phase using the deformation of the Fermi surface. Such deformation of
the Fermi surface can be constructed by an in-plane Zeeman field and SO
coupling. The center of the Fermi surface is also shifted due to the Zeeman
field, and as a consequence, the BCS type of pairing always becomes
challenging even with a small Zeeman field. A very large FFLO phase can be
observed in the parameter space. The deformation of the Fermi surface breaks
the rotational symmetry of the system, thus can create the
FFLO phase with a unique FFLO momentum $\mathbf{Q}$.}
\label{fig-idea}
\end{figure}

The symmetry breaking has a direct consequence on the formation
of FFLO superfluids. Before the presentation of our numerical results, we
first illustrate the basic physical picture for the formation of FFLO
superfluids. For the Fermi gas with only Zeeman field, see Fig. \ref%
{fig-idea}a, the two mismatched Fermi surfaces always form concentric
circles, therefore for the $s$-wave pairing, the up- and down-spins acquire
different Fermi momentum, \textit{i.e.}, $\mathbf{k}+\mathbf{Q}/2,\Uparrow $
and $-\mathbf{k}+\mathbf{Q}/2\Downarrow $, with $\Uparrow $ and $\Downarrow $
spins in the pseudospin representation and $\mathbf{Q}$ as the total
momentum of the Cooper pairs. The free energy of the system satisfies the
following basic property,
\begin{equation}
F(\mathbf{Q})=F(|\mathbf{Q}|).  \label{eq-F}
\end{equation}%
Here the Zeeman field only fixes the direction of the spin, but does not fix the direction
of the momentum axis, therefore the free energy should be invariant under rotation
of the momentum $\mathbf{Q}$. Mathematically, it can also be understood from the fact that
the total free energy depends on $\mathbf{k}^2$ , $\mathbf{Q}^2$ and $\mathbf{k}\cdot\mathbf{Q}$,
thus the summation over $\mathbf{k}$ should
be independent of the direction of $\mathbf{Q}$, see also Ref. \onlinecite{Sheehy} for more details. Physically,
Eq. \ref{eq-F} means that the total momentum of the Cooper pair can take any direction
by spontaneous symmetry breaking, therefore the ground state FFLO phase is infinity-
fold degenerate. Generally in the numerical simulation, we artificially set $\mathbf{Q}$ along a
particular direction and demonstrate that the FFLO phase indeed has a lower energy
than the regular BCS superfluid ($\mathbf{Q}$ = 0). Due to the Pauli paramagnetic depairing
effect, the FFLO phase only survives in a very narrow parameter regime, see also the
numerical results in Fig. \ref{fig-phase}a. In realistic system, any weak scattering induced
by disorder effect can lead weak coupling between the degenerate ground states manifold, making the LO superfluids,
which can be regarded as a superposition of the two FF superfluids with total momentum $\mathbf{Q}$ and $-\mathbf{Q}$,
as the true ground states, and the LO superfluids still respect the basic symmetry arguement in Eq. \ref{eq-F}.

The physical picture is totally different when the SO coupling is presented,
as schematically shown in Fig. \ref{fig-idea}b. In this case the Fermi
surface is deformed and the center of the Fermi surface is no longer located
at $\mathbf{k}=0$, therefore breaks the inversion symmetry.
Here we should notice that the deformation of the Fermi
surface depends strongly on the direction of the SO coupling and Zeeman
field. For the model we consider here, the deformation is along the $y$
direction. In the pseudospin representation (the eigenstates of single
particle Hamiltonian), we have both singlet pairing and triplet pairing,
where the triplet pairing will not be destroyed by strong Zeeman field, thus
the FFLO phase can be observed in a much larger parameter regime.
The deformation of the Fermi surface makes the FFLO phase always energetically favorable even with
a small Zeeman field. In our numerics, we find that the FFLO vector $\mathbf{Q}$ is
along the deformation direction of the Fermi surface. The inversion symmetry breaking
directly lead to $F (\mathbf{Q}) \neq F (-\mathbf{Q})$, which stabilize the FF superfluids against the
formation of LO superfluids phase.

Generally, the mismatch of the Fermi surface is the basic route to the FFLO
phase, and such mismatched Fermi surface can be created by population
imbalance \cite{Zwierlein06a,Zwierlein06b, Hulet1}, Zeeman field \cite{Hu}
or mass imbalance \cite{He06, Conduit}. In this work, together with our
previous work \cite{Zhen}, we demonstrate that the FFLO phase can be created
more efficiently through the deformation of the Fermi surface, which can be
constructed by SO coupling, or, non-Abelian gauge field \cite{Ruseckas,Xiongjun,xiongjun2,CWZhang,Campbell,Jay, non-Abelian}, and Zeeman
field. Notice that the generation of non-Abelian gauge fields is a subject
of intensive investigations in ultracold atoms in the past decade, see a
recent review in Ref. \onlinecite{non-Abelian}. For this new route, the
Zeeman field is still needed. Otherwise the system has the time-reversal
symmetry and the band structure should satisfy $E_{\mathbf{k}\uparrow }=E_{-%
\mathbf{k},\downarrow }$, which means that two Fermions with opposite
momentum on the Fermi surface can always form BCS Cooper pairs efficiently (the pairing is not necessary
in the singlet channel), leading to BCS superfluids, instead of FFLO phases. Our route
here, however, shows that the FFLO phase may be observed even with small
Zeeman field (thus small population imbalance). It therefore represents a new
driven mechanism for FFLO superfluid.

\section{Numerical Details}

\label{sec-numericaldetails}

The order parameter $\Delta $, chemical potential $\mu $, and the FFLO
momentum $\mathbf{Q}$ should be solved self-consistently due to the
conservation of atom number, i.e.,
\begin{equation}
{\frac{\partial \Omega }{\partial \mu }}=-n,\quad {\frac{\partial \Omega }{%
\partial \Delta }}=0,\quad {\frac{\partial \Omega }{\partial \mathbf{Q}}}=0.
\label{eq-eq3}
\end{equation}%
Here $\Delta $ and $\mathbf{Q}$ are used to minimize the thermodynamical
potential $\Omega $. We consider three different quantum phases: the normal
phase with $\Delta =0$ and $\mathbf{Q}=0$ (for the normal gas $\mathbf{Q}$
does not enter the effective free energy, thus can be any value. We force $%
\mathbf{Q}=0$); The BCS-type of superfluid with $\mathbf{Q}=0$ but $\Delta
\neq 0$; and the FFLO phase with $\mathbf{Q}\neq 0$ and $\Delta \neq 0$.
When we fix $\mathbf{Q}=0$ then only BCS type of superfluid phase and the
normal gas can be obtained. Throughout this work, we use two different
strategies to check the influence of $\mathbf{Q}$ on the formation of the
FFLO phase. In the first strategy, we enforce $\mathbf{Q}=0$ while in the
other strategy, we let $\mathbf{Q}$ as a free parameter. For the results at $%
\mathbf{Q}\neq 0$, these two strategies yield the energy difference between
the FFLO ground state and the possible BCS superfluid excited state, which
is crucial for the stability of the FFLO phase at finite temperature.
Throughout this work, $Q_c = 10^{-3} K_F$ is used.

Because the Zeeman field is applied along the $x$-axis, the population
imbalance should be defined using the eigenstates of $\sigma _{x}$, instead
of $\sigma _{z}$. Since $\langle \sigma _{x}\rangle =\sum_{\mathbf{k}%
}\langle c_{\mathbf{k},\uparrow }^{\dagger }c_{\mathbf{k},\downarrow }+\text{%
h.c}\rangle $, we have
\begin{equation}
P={\frac{\langle \sigma _{x}\rangle }{n}} =
\frac{\sum_{\mathbf{k}}\langle c_{\mathbf{k},\uparrow }^{\dagger }c_{\mathbf{k},\downarrow }+\text{h.c}\rangle}{n}=
{\frac{1}{2n}}\sum_{\mathbf{k}%
,\lambda }\psi _{\mathbf{k},\lambda }^{\dagger }%
\begin{pmatrix}
\sigma _{x} & 0_{2\times 2} \\
0_{2\times 2} & -\sigma _{x}%
\end{pmatrix}%
\psi _{\mathbf{k},\lambda }.
\label{eq-population}
\end{equation}%
Here $\psi _{\mathbf{k},\lambda }$ is the eigenstate of the
effective Hamiltonian $H_{\text{eff}}$, i.e., $H_{\text{eff}}\psi _{\mathbf{k%
},\lambda }=E_{\lambda }\psi _{\mathbf{k},\lambda }$.

In our calculation, we choose the energy unit as the Fermi energy $E_{F}$ of
the system without interaction, Zeeman field and SO coupling. The
corresponding length scale $K_{F}^{-1}$ is defined through the Fermi
momentum $K_{F}$. At finite temperature, the 2D system does not have the
long-range order due to the phase fluctuation and the relevant physics is
the Kosterlitz-Thouless transition\cite{Gong12}. In this paper, we restrict
to the physics at zero temperature, where the mean-field theory is still valid.
For this specific model, we find ${\bf Q} = (0, Q)$, which means that the
FFLO momentum is along the Fermi surface direction, see Fig. \ref{fig-idea}.
We notice that the direction of the FFLO vector $\mathbf{Q}$ is also consistent with the results
in solid state systems with weak SO coupling, see Ref. \onlinecite{Vitctor}.

\section{Phase diagram}

\label{sec-phasediagram}

\begin{figure}[tbp]
\centering
\includegraphics[width=5in]{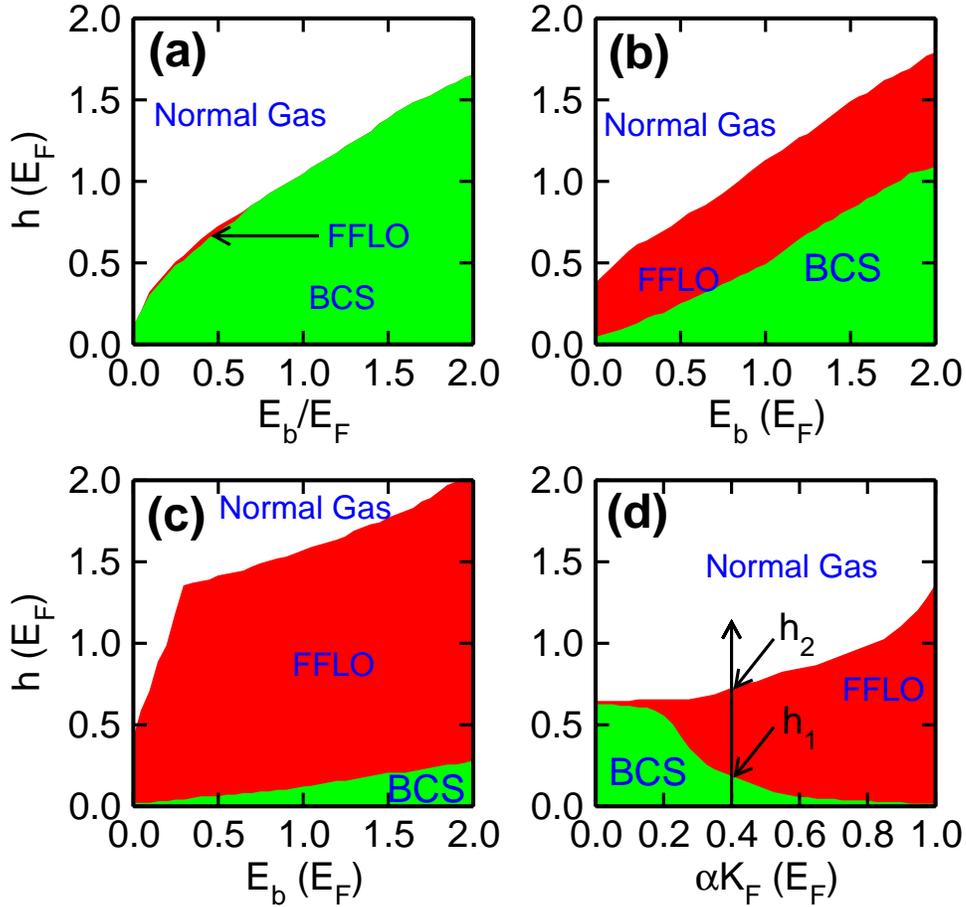}
\caption{(Color online). Phase diagram of the 2D degenerate Fermi gas in the
presence of Rashba-type SO coupling and an in-plane Zeeman field. (a)
results with vanishing SO coupling. (b), (c) correspond to the results with
SO coupling $\protect\alpha K_{F}=0.5$ and $\protect\alpha K_{F}=1.0$,
respectively. (d) shows the phase diagram in the $h-\protect\alpha K_{F}$
plane at $E_{b}=0.4E_{F}$. $h_{1}$ ($h_{2}$) defines the boundary between
BCS (FFLO) and FFLO (normal gas) phases.}
\label{fig-phase}
\end{figure}

\begin{figure}[tbp]
\centering
\includegraphics[width=5in]{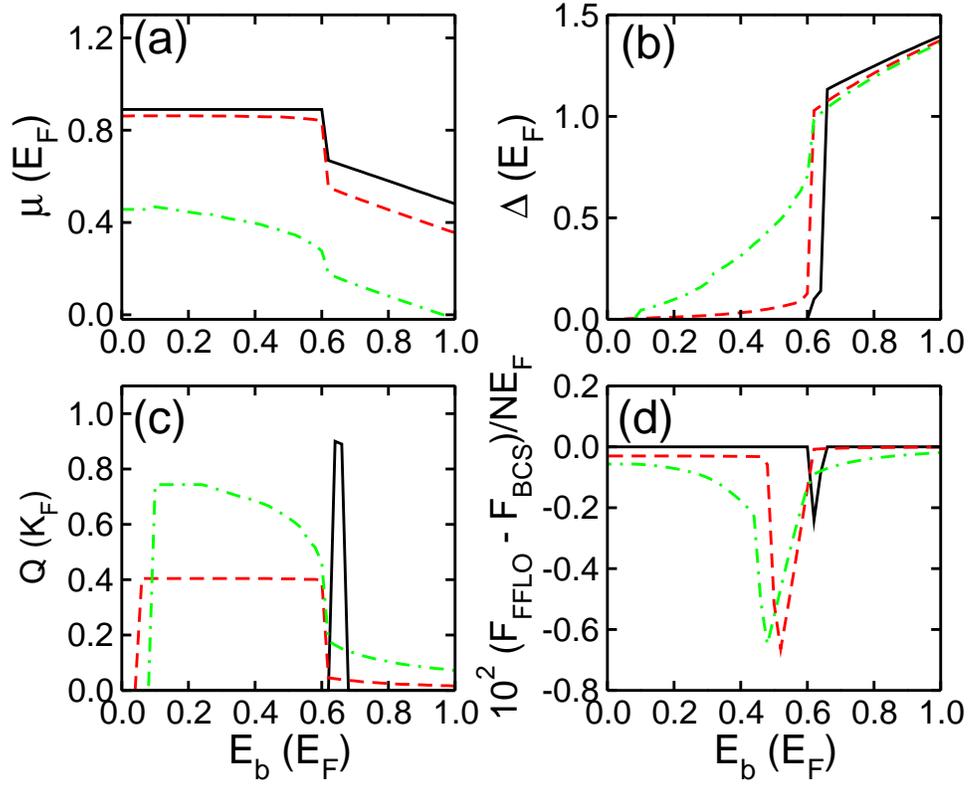}
\caption{(Color online). Evolution of chemical potential (a), order
parameter (b), and FFLO vector $\mathbf{Q}$ (c) as a function of the binding
energy. In (d) we plot $(F_{\text{FFLO}}-F_{\text{BCS}})/nE_{F}$ \textit{vs.}
$E_{b}$. In all calculations we set $h=0.8E_{F}$. The solid line, dashed
line and dash-dotted line correspond to $\protect\alpha K_F = 0.0$, 0.5 and
1.0, respectively.}
\label{fig-becbcs}
\end{figure}

\begin{figure}[tbp]
\centering
\includegraphics[width=5in]{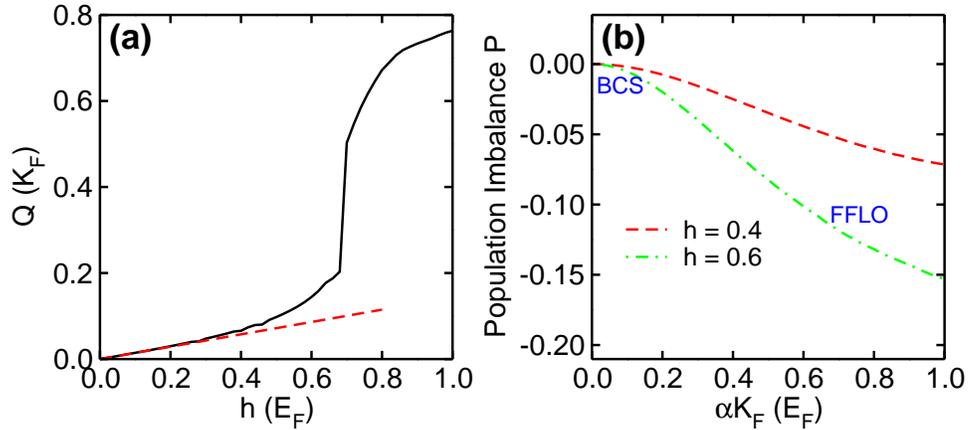}
\caption{(a) FFLO momentum $Q$ as a function of Zeeman field (solid line), and the
dashed line is the linear fitting at small Zeeman field regime, which give Q = 0.1434h.
(b) Population imbalance P (Eq. \ref{eq-population}) as a function of SO coupling strength. The
parameters are: (a), $E_b = 0.4E_F$ , $\alpha K_F = 1.0E_F$; (b) $E_b = 0.4E_F$ .}
\label{fig-sigmax}
\end{figure}

\begin{figure*}[tbp]
\centering
\includegraphics[width=0.98\textwidth]{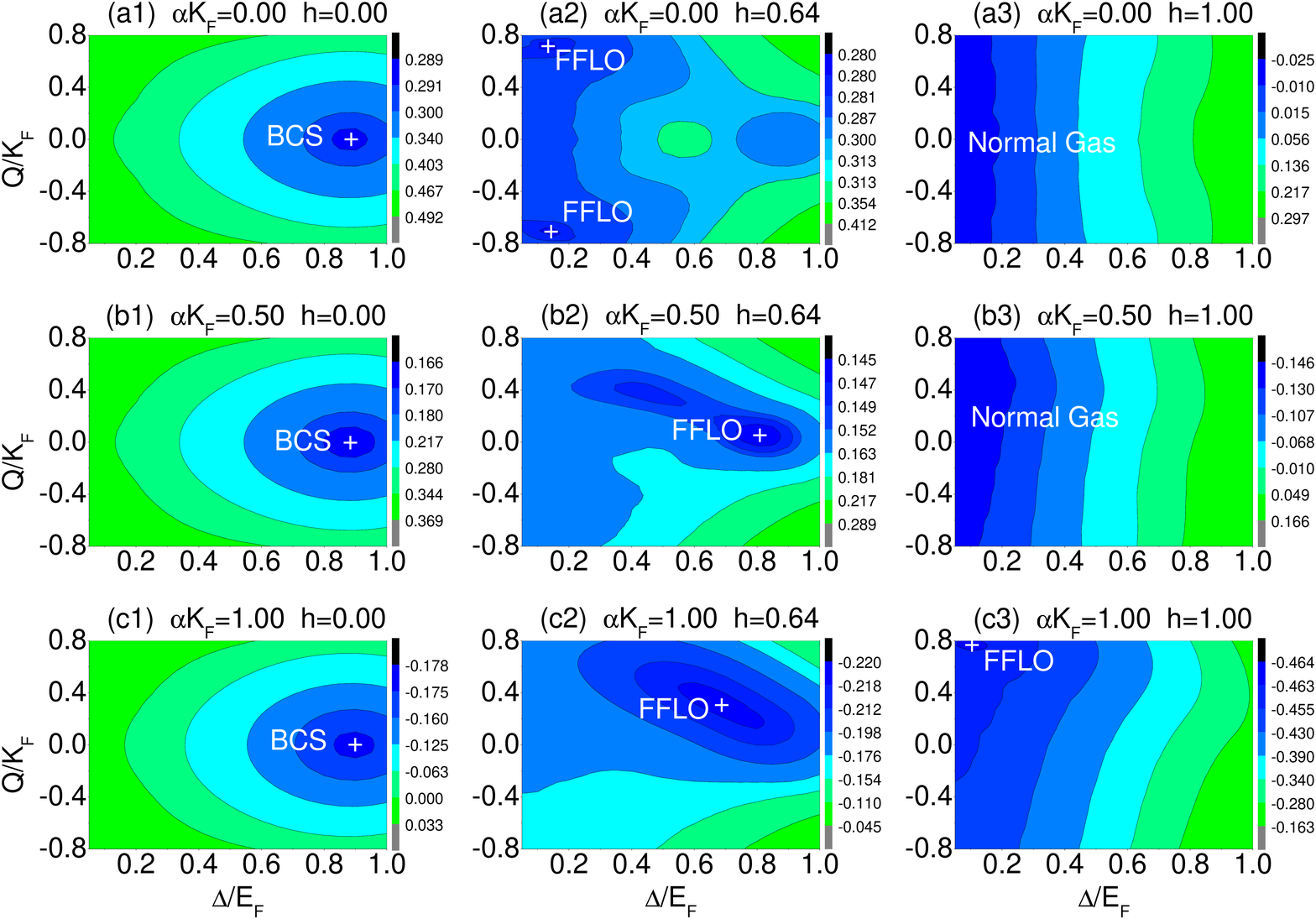}
\caption{(Color online). Influence of SO coupling and in-plane Zeeman field
on the free energy per particle, $F/(nE_{F})$, in the $Q-\Delta $ plane. The
cross symbol in each panel corresponds to the self-consistent solution of
Eq. \protect\ref{eq-eq3}. In (a3) and (b3) only normal gas can be observed,
thus $\Delta =0$, and $Q$ can be any value because it doesn't enter the free
energy.}
\label{fig-contour}
\end{figure*}

We first present the phase diagram with different SO coupling strength and
Zeeman field in Fig. \ref{fig-phase}. Without SO coupling, see Fig. \ref%
{fig-phase}a, we see that the FFLO phase only exists in an extremely narrow
parameter regime. When $E_{b}\geq 0.7E_{F}$, the FFLO phase disappears, thus
such a phase can be only observed in the weak binding energy regime, for
instance, $E_{b}\in (0.15,0.7)E_{F}$. Similarly, the FFLO phase can also be
observed in the 3D system, see Ref. \onlinecite{Zhen}; however, the FFLO
phase in 3D Fermi gases can only be observed near the unitary regime within
a small parameter region, and the small FFLO regime can be easily missed out
in realistic experiments, which is also one of the main reasons why the FFLO
phases cannot be observed in recent experiments in 3D Fermi gases \cite%
{Zwierlein06a, Zwierlein06b, Hulet1}. With an increasing SO coupling
strength, see Fig. \ref{fig-phase}b for $\alpha K_{F}=0.5E_{F}$ and Fig. \ref%
{fig-phase}c for $\alpha K_{F}=1.0E_{F}$, we find that the FFLO phase regime
is greatly enlarged. In the strong SO regime in Fig. \ref{fig-phase}c, we
even observe that the phase diagram is almost fully filled by the FFLO
phase, while the BCS superfluid phase is greatly suppressed and only
survives in a very small regime. To see the impact of SO coupling more
clearly, we plot in Fig. \ref{fig-phase}d the phase diagram in the $h-\alpha
K_{F}$ plane with $E_{b}=0.4E_{F}$. We define the boundary between BCS
superfluid and FFLO phase as $h_{1}$ and the boundary between FFLO phase and
normal gas as $h_{2}$ for convenience, see Fig. \ref{fig-phase}d. We observe
$h_{1}$ decreases while $h_{2}$ increases with the increasing SO coupling
strength, therefore the FFLO phase is greatly enlarged in the strong SO
coupling regime. It should be noticed that in 3D Fermi gases $h_{2}$
slightly decreases with the increasing SO coupling strength \cite{Zhen}. In
the strong SO coupling region, $h_{1}$ becomes very small, but never becomes
zero because the the Zeeman field is essential for the FFLO phase, which
breaks the time-reversal symmetry.

We plot the evolution of chemical potential, order parameter and $Q$ as a
function of binding energy in Fig. \ref{fig-becbcs}, where the Zeeman field
is fixed to $h=0.8E_{F}$. As we decreases the binding energy, we observe a
sudden drop of the order parameter in Fig. \ref{fig-becbcs}b at zero SO
coupling strength due to the Pauli paramagnetic depairing effect, following
which there is a small regime that supports FFLO phase, see also the solid
line in Fig. \ref{fig-becbcs}c, $Q\neq 0$. With the increasing SO coupling
strength, we see that the change of $\Delta $ becomes a smooth function of $%
E_{b}$, and in a much larger parameter regime we can observe the FFLO phase
with non-zero $Q$. The results in Fig. \ref{fig-becbcs}c clearly demonstrate
the enlargement of FFLO superfluid phases observed in Fig. \ref{fig-phase}.

The FFLO superfluids in our model may be directly observed at finite temperature.
We denote $F_{\text{FFLO}}$ as the free energy obtained by letting $\mathbf{Q}$ as a free parameter, while $F_\text{BCS}$
as the free energy by enforcing $\mathbf{Q} = 0$. In the FFLO phase regime, $F_{\text{BCS}}$ represents the
free energy of BCS excited states, therefore the energy difference per particle between
$F_\text{FFLO}$ and $F_\text{BCS}$ , i.e.,
\begin{equation}
\delta F = (F_{\text{FFLO}}-F_{\text{BCS}})/nE_F,
\label{eq-deltaF}
\end{equation}
which directly characterizes the stability of the FFLO phase (i.e., the larger $|\delta F |$, the
more stable FFLO phase). Obviously, when $\mathbf{Q} = 0$, $\delta F = 0$. The numerical results
are presented in Fig. \ref{fig-becbcs}d, where we clearly see the enhancement of $\delta F$ due to the
SO coupling. However, in 2D Fermi gases the enhanced factor is about two order of
magnitude smaller than that in SO coupled 3D Fermi gases.

\begin{figure}[tbp]
\centering
\includegraphics[width=5in]{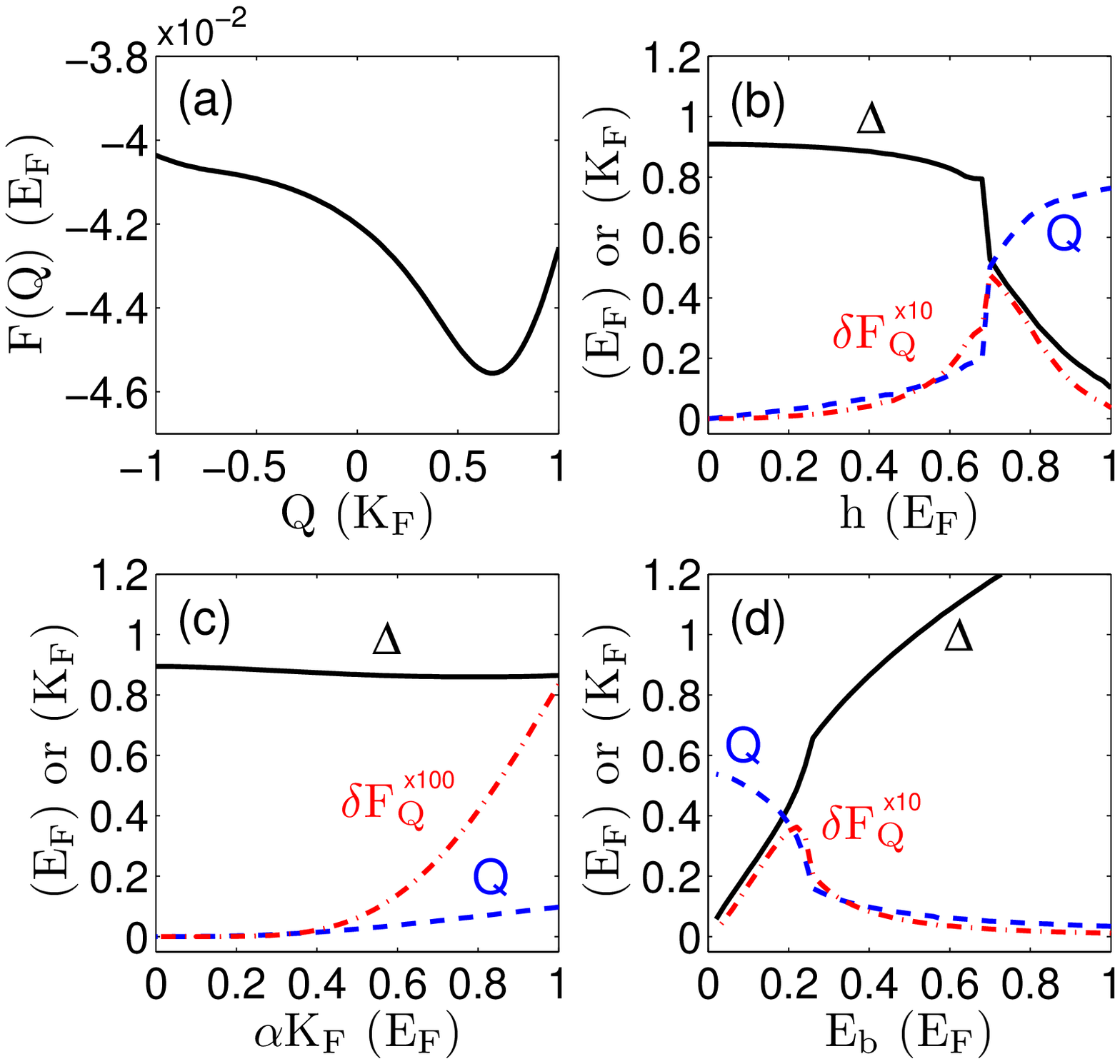}
\caption{(a) Free energy as a function of Q, where $F (Q) \neq F (-Q)$ stabilize the
FF superfluids against the formation of LO superfluids. (b-d) $(F (-Q) - F (Q))/nE_F$,
$\Delta/E_F$ and $Q/K_F$ as a function of SO coupling, Zeeman field and binding energy,
respectively. The parameters used in all figures are: (a): $E_b = 0.4E_F$, $\alpha K_F = 1.0$,
$h = 0.8E_F$; (b) $E_b = 0.4E_F$, $\alpha K_F = 1.0$; (c) $E_b = 0.4E_F$ , $h = 0.5E_F$; and (d)
$\alpha K_F = 1.0E_F$, $h = 0.5E_F$.
}
\label{fig-deltaF}
\end{figure}

\begin{figure}[tbp]
\centering
\includegraphics[width=5in]{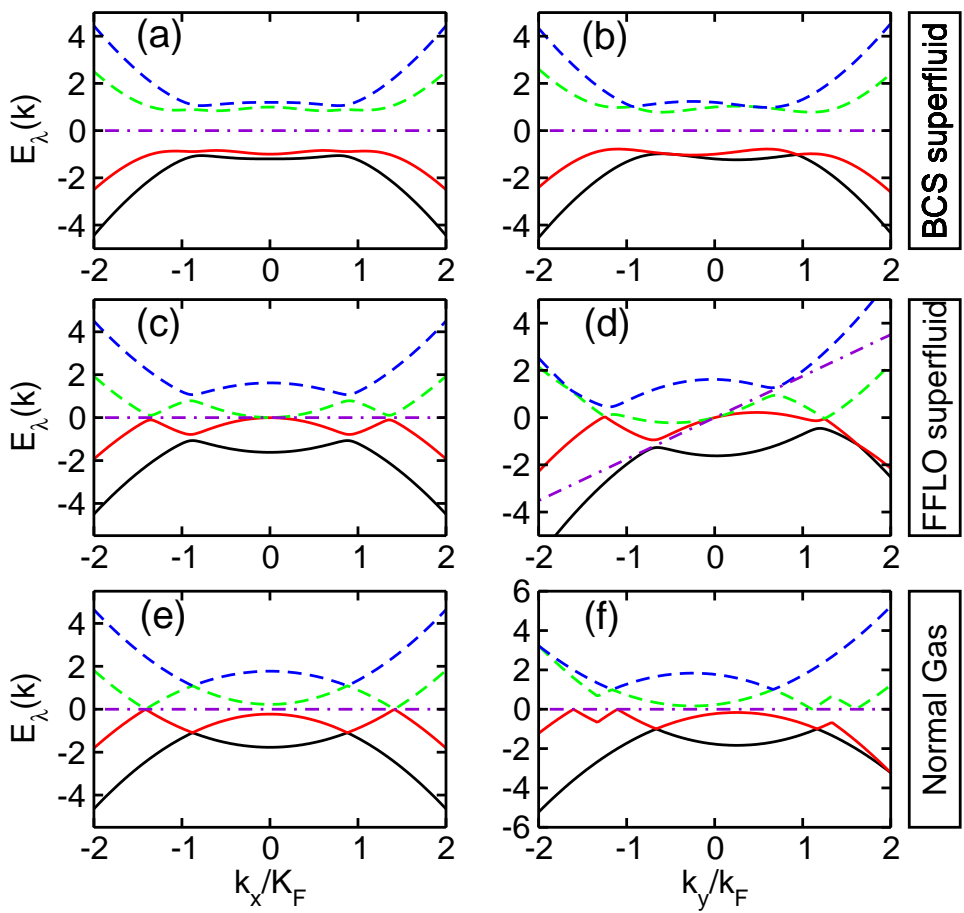}
\caption{(color online). Eigenvalues $E_{\protect\lambda }$ ($\protect%
\lambda =1$, 2, 3, 4) of the SO coupled degenerate Fermi gas. (a), (b)
correspond to the typical eigenvalues $E_{\protect\lambda }$ for the BCS
superfluid with parameters $h=0.2$, $E_{b}=0.4$, $\protect\alpha k_{F}=1.0$.
(c), (d) correspond to the typical eigenvalues $E_{\protect\lambda }$ for
the FFLO superfluid with parameters $h=0.4$, $E_{b}=0.4$, $\protect\alpha %
k_{F}=1.0$. (e), (f) correspond to the typical eigenvalues $E_{\protect%
\lambda }$ for the normal gas with parameters $h=1.0$, $E_{b}=0.4$, $\protect%
\alpha k_{F}=1.0$. The first column shows the results along the $x$
direction, while the second column shows the results along the $y$
direction. The energies are in unit of $E_F$. In each panel, the dash-dotted
line represent Tr($H_{\text{eff}}({\bf k})$), see Eq. \ref{eq-elambda}.}
\label{fig-Ek}
\end{figure}

\begin{figure}[tbp]
\centering
\includegraphics[width=5in]{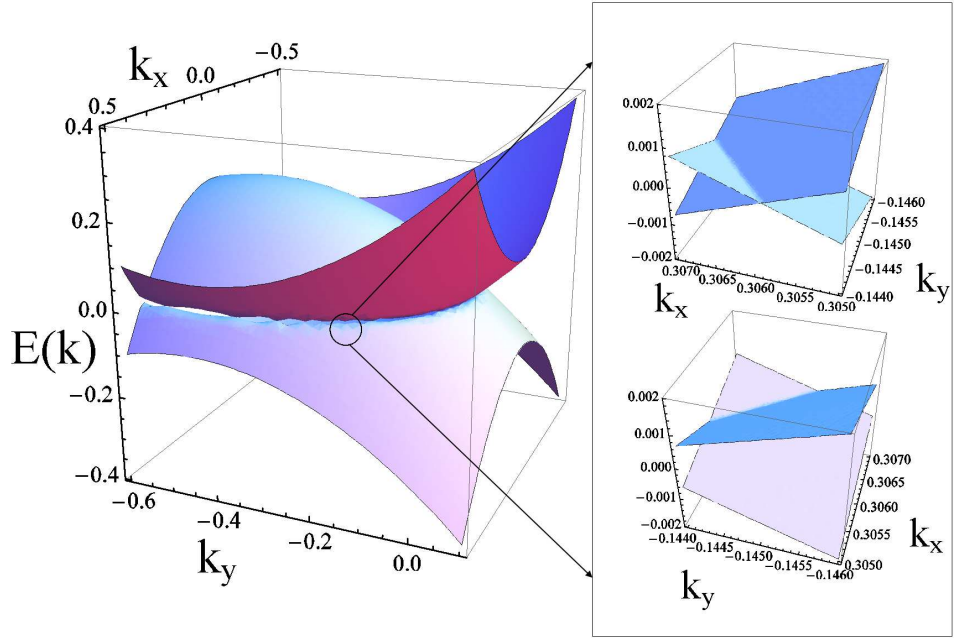}
\caption{Gapless excitations for the FFLO phase. Near $E_{\mathbf{k},\protect%
\lambda }=0$, the energy shows a clear linear dispersion. $k_x$ and $k_y$
are in unit of $K_F$. In 2D system the linear dispersion is essential to
make the FFLO phase robust against low-energy fluctuations.}
\label{fig-linear}
\end{figure}

\begin{figure}[tbp]
\centering
\includegraphics[width=5in]{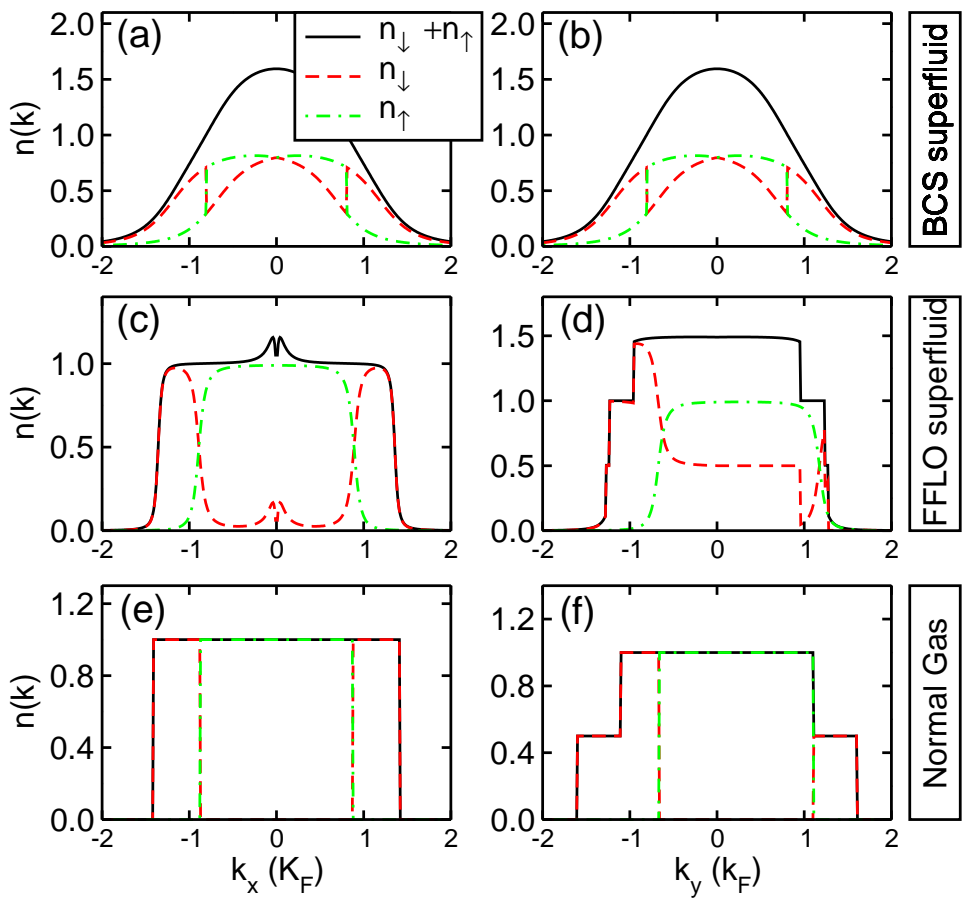}
\caption{(color online). Momentum distributions $n_{\protect\sigma %
}(k)=\langle c_{\protect\sigma }^{\dagger }(k)c_{\protect\sigma }(k)\rangle $
and $n=n_{\downarrow }+n_{\uparrow }$ for different quantum phases. Other
parameters are exactly the same as that in Fig. \protect\ref{fig-Ek}.}
\label{fig-nk}
\end{figure}

In Fig. \ref{fig-becbcs}b, we see that in the BCS superfluid regime ($E_{b}>0.7E_{F}$), the order
parameter decreases with the increasing SO coupling, which is in sharp contrast to that
for SO coupled BEC-BCS crossover with Z direction Zeeman field. Generally, with
the $Z$ direction Zeeman field, the SO coupling plays the role of increasing the density
of states near the Fermi surface, which increases the order parameter as well as the
critical temperature. With an in-plane Zeeman field, the SO coupling plays a totally
different role. Firstly, the in-plane Zeeman field deforms the Fermi surface, thus any
small deformations leads to small finite momentum $Q$, as shown in Fig. \ref{fig-sigmax}a. In the small
Zeeman field regime, the momentum $Q\propto h$, while in the large Zeeman field regime, it
become a nonlinear behavior. Secondly, the SO coupling can enhance the population
imbalance, see Fig. \ref{fig-sigmax}b, thus renders the decrease of the order parameter as observed in
Fig. \ref{fig-becbcs}b. In the FFLO phase regime, the order parameter increases with the increasing
SO coupling strength due to the formation of the FFLO phase. Notice that in our
model, the FFLO superfluid can appear with extremely small population imbalance,
thus it is driven by the interplay between the deformation of Fermi surface and the
superconducting order, instead of the original idea of FFLO superfluid which arises
from the interplay between magnetism and superconducting order. The new driven
mechanism represents a more efficient way to create the FFLO superfluid.

\section{Free Energy Landscape}

\label{sec-Flandscape}

To understand the results more clearly, we plot the free energy per particle in the $Q-\Delta$
plane, where the global minimum of the free energy marked by the cross symbol in each
panel corresponds to the self-consistent solution of Eq. \ref{eq-eq3}. The left to right shows the
nfluence of the SO coupling on the formation of the FFLO phase, while the top to
down shows the influence of the Zeeman field. Note that without SO coupling, the free
energy is a symmetric function of Q, therefore in Fig. \ref{fig-contour}a2 there are two degenerate
FFLO ground states at $\pm Q$ because we have assumed $\mathbf{Q}$ is along the $y$ direction in
our numerical simulation. This is not true in the real system where the free energy
only depends on the magnitude of $|\mathbf{Q}|$ due to the rotation symmetry, see Eq. \ref{eq-F}. The
free energy is still a symmetric function with respect to $Q$ when $h = 0$. However it
becomes an asymmetric function when both Zeeman field and SO coupling strength
become non-zero, thus only one global minimum can be found in the $Q-\Delta$ plane in
Figs. \ref{fig-contour}b2, c2, c3, and the ground state is unique (hence $\mathbf{Q}$ is unique). The plot of the
free energy in the $Q-\Delta$ plane directly reflects the effect of the rotational symmetry
breaking of the effective Hamiltonian. With the increasing Zeeman field, the order
parameter decreases due to the formation of the FFLO phase. Due to the increases of
$h_2$ , the boundary between FFLO phase and normal gas, we observe the FFLO phase in
the strong Zeeman field and strong SO coupling region in Fig. \ref{fig-contour}c3.

It is well known in solid materials that the LO superfluids, which is the superposition
of FF superfluids with total momentum $\mathbf{Q}$ and $-\mathbf{Q}$, is more energetically favorable in
realistic systems. The basic reason is that the deformation of Fermi surface is very small,
thus $F(\mathbf{Q})\approx F(-\mathbf{Q})$. As a results, the coupling between  FF superfluids with momentum $\mathbf{Q}$ and
$-\mathbf{Q}$ lead to the formation of LO superfluids with slightly lower energy. The coupling between different FF superfluids
is even significant in the degenerate ground states manifolds. So it means that the energy difference between $F({\bf Q})$ and
$F(-\mathbf{Q})$, where $\mathbf{Q}$ is the FFLO superfluids momentum obtained using the previous
procedures, can be used to qualify whether FF superfluids is more stable than the
LO superfluids. Hence we define
\begin{equation}
\delta F_Q = (F(-\mathbf{Q}) - F(\mathbf{Q}))/nE_F.
\end{equation}
Obviously, when $\mathbf{Q} = 0$, $\delta F_Q = 0$. On the other hand, $\delta F_Q = 0$ when $\Delta = 0$. Note that
$\delta F$ defined in Eq. \ref{eq-deltaF} and $\delta F_Q$ defined above have totally different physical meanings,
and should not be mixed up. In Fig. \ref{fig-deltaF}a, we show that the breaking of inversion
symmetry readers $F (\mathbf{Q}) \neq F (-\mathbf{Q})$. We also plot $\delta F_Q$ , $\Delta/E_F$ and $Q/K_F$ as a function
of SO coupling, Zeeman field and binding energy in Fig. \ref{fig-deltaF}c-d. We see that the increase
of SO coupling monotonically increase $Q$ and hence $\delta F_Q$ also increase monotonically.
However, when $\Delta$ or $Q$ has a sudden jump at some point, see Fig. \ref{fig-deltaF}b, d, then $\delta F_Q$ may
take a maximum at these points. In the condition with strong Zeeman field or large
binding energy, the FFLO phase is suppressed, therefore $\delta F_Q$ approaches zero as the
increase of these parameters. Here the most interesting result we observed is that $\delta F_Q$
can be as large as 0.1, and this large energy difference ensures that the FF superfluid
phase has much lower energy than the LO superfluid phase. Similar result can not be observed in solid materials. 

\section{Measurement of the FFLO phase}

\label{sec-measurement}

The three different phases have different properties which can be used for
the identification of these phases. In Fig. \ref{fig-Ek}, we plot the
typical band structures $E_{\lambda }$, $\lambda =1$, 2, 3, 4, for the BCS
superfluid, the FFLO phase and the normal gas. Due to the rotational
symmetry breaking, we have to plot the dispersions along the $k_{x}$ and $%
k_{y}$ axes, respectively. For a typical BCS superfluid ($\mathbf{Q}=0$) in
Fig. \ref{fig-Ek}a and Fig. \ref{fig-Ek}b, we see that the system is always
gapped and the band structure is always symmetric about $\mathbf{k}=0$ for
the dispersion along $k_{x}$. While along the $k_{y}$ axis, such symmetry is
absent. In fact we can verify exactly that the BCS superfluid is always
gapped. However for the FFLO phase, the superfluid becomes gapless along
both $k_{x}$ and $k_{y}$ axes. Along the $k_{x}$ axis the band structure is
symmetric about $\mathbf{k}=0$, but along the $k_{y}$ direction such
symmetry is broken. For the FFLO phase we observe $\sum_{\lambda }E_{\lambda
}\neq 0$ because $\mathbf{Q}\neq 0$ (see numerical results in Fig. \ref{fig-Ek}),
which is consistent with our symmetry
analysis in sec. \ref{sec-FSsymmetry}. Note that the gapless excitation is a
typical feature of the FFLO phase, as have been pointed out in literature
\cite{Sheehy}. In the vicinity of the gapless excitation, see Fig. \ref%
{fig-linear}, the dispersion becomes linear which is essential to ensure
that the FFLO phase is robust against the low-energy fluctuations. Here we
should emphasize that not all FFLO phases are gapless. The FFLO state may
become gapless only when $Q$ is relatively large, while for a small $Q$
(near the boundary between FFLO and BCS superfluid) the FFLO phase is still
gapped, similar to that in the BCS superfluid. For the normal gas the band
structure also shows strong deformation along the $k_{y}$ axis, as seen in
Fig. \ref{fig-Ek}e and Fig. \ref{fig-Ek}f.

The corresponding momentum distributions $n_{\sigma }=\langle c_{\textbf{k}\sigma}^{\dagger }c_{\textbf{k}\sigma }\rangle $ and $n=n_{\uparrow }+n_{\downarrow }$
provide an important tool to detect the properties of the FFLO state because
they can be directly measured via free expansion of the atomic cloud. We
plot the momentum distributions in Fig. \ref{fig-nk} for three different
phases presented in Fig. \ref{fig-Ek} at zero temperature. The dispersion
properties of the band structures can be directly reflected on the
corresponding momentum distributions. We see that for three different
quantum phases, the momentum distributions are always symmetric about $%
\mathbf{k}=0$ along the $k_{x}$ direction, while show strong asymmetric
along the $k_{y}$ direction. However, the sum of the momentum distributions $%
n$ for spin up and spin down components still shows perfect symmetry about $%
\mathbf{k}=0$ along both $k_{x}$ and $k_{y}$ directions. Therefore detecting
the asymmetry of the superfluid is not sufficient for the identification of
the FFLO phase. To identify the superfluid nature of the FFLO phase, we have
to rotate the sample to create vortices, which is a direct evidence of
superfluidity. Near the boundary between difference phases, the fluctuation
effect may become significant thus the phase boundary region is not suitable
for the observation of vortices. With the large FFLO phase region in our
model we can safely choose some parameters in the middle of the FFLO phase
region where the fluctuation effect should be minimized. The large FFLO
superfluid phase ensures that it will not be missed out in future realistic
experiments.

\begin{figure}[tbp]
\centering
\includegraphics[width=5in]{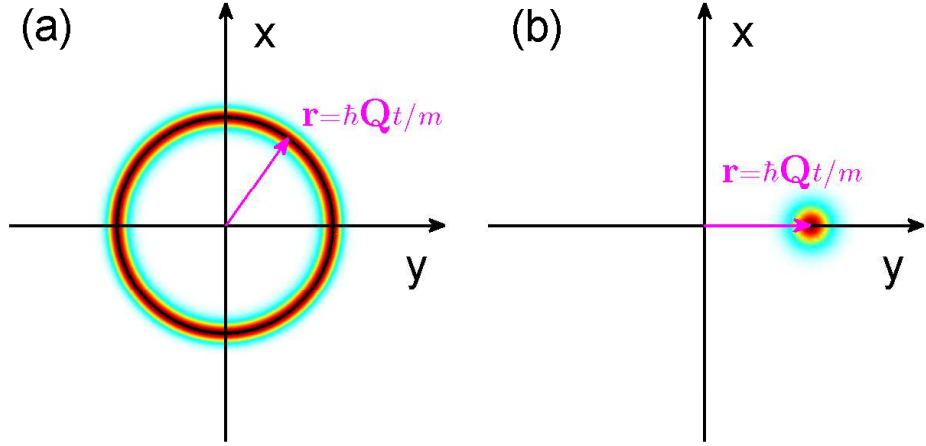}
\caption{Typical time-of-flight image for the degenerate Fermi gas with (a) and
without (b) inversion symmetry. For the system without inversion symmetry, the
FFLO momentum $\mathbf{Q}$ is along the principle Fermi surface deformation direction, and it
can be directly measured in experiments}
\label{fig-measurement}
\end{figure}

The properties of the FFLO phase may be measured using a number of methods
developed in ultracold atom systems, for instance, shot-noise correlation
\cite{Altman} and density-density correlation measurement \cite{Koponen,
Koponen2}, which shows a peak at the Cooper pair momentum $\mathbf{Q}$.
After released from a trapping potential, the free expansion of the Fermi
cloud has a peak at $\mathbf{r}=\hbar \mathbf{Q}t/m$, therefore the direct
measurement of the FFLO momentum $\mathbf{Q}$ is possible \cite{Stein}. In
our model when $\mathbf{Q}$ is unique, repeated measurement to determine the
FFLO momentum becomes possible,
see Fig. \ref{fig-measurement}.. In the FFLO phase without SO coupling, the
ground state is independent of the direction of $\mathbf{Q}$,
thus only a circle with radius $\hbar|\mathbf{Q}|t/m$ can be observed,
see Fig. \ref{fig-measurement}. So the time-of-flight imaging
provides the most convenient way to probe the symmetry effect of the degenerate
Fermi gas. In other words, the time-of-flight imaging directly reflect the deformation
direction of the Fermi surface. The FFLO phase can also be measured using
the Fourier sampling of time-of-flight images proposed by Duan \cite{Duan}.
The gapless excitations in the FFLO phase may be observed using the Bragg
spectroscopy \cite{Stein}.

\section{Comparison between cold atom and solid materials}

\label{sec-vssolids}

We notice that the similar model (Rashba SO coupling, in-plane Zeeman field, {\it etc.}) has been discussed in condensed
matter physics in the context of noncentrosymmetric superconductors\cite{Yip02, Agterberg, Dim, Kaur, Matsunaga08, Oka06, Yuan, Yokoyama, Lee, Tada, Frigeri},
and our observation that ${\bf Q}$ perpendicular to the direction of Zeeman field as well as the SO interaction significantly
broadens the FFLO phase in parameter space are consistent with that in solids\cite{Vitctor}. It does not means that the physics in solids and
cold atoms are similar or identical. In the following, we summarize some of the notable differences between these two totally different systems.
Some of these differences have been briefly discussed in our previous work\cite{Zhen}.

\subsection{Different driving mechanism for FFLO phases}

The driving mechanisms for FFLO phases in solid state systems and cold atom systems are quite different. In solid state systems,
Zeeman field and SO energy are generally much smaller than the Fermi energy, the asymmetry of the Fermi surface is still very small,
and the FFLO phase is mainly induced by the interplay between magnetic and superconducting order. In our
work\cite{Zhen}, we propose a totally different route for the creation of FFLO phase. In cold atom system, the Fermi energy, SO coupling
energy and Zeeman field energy are at the same order, the deformation of the Fermi surface becomes
coupling energy at the order of the Fermi energy, the deformation of the Fermi surface becomes significant, therefore the FFLO phase can
still be observed even for system with small population imbalance. In this new mechanism, the FFLO phase is induced  by the interplaying between
asymmetry of Fermi surface and superconducting order. Note that in Ref.\onlinecite{Patrick}, the ratio between SO coupling energy and Fermi energy
is of the order of $0.1 - 0.5$ in the SrTiO$_3$/LaAlO$_3$ oxide interface, and we expect this new mechanism applies to this solid material. Here we
need to emphasize that the basic mechanism for finite momentum pairing in some of the solid materials are still not well theoretically
understood due to the complicated spin fluctuating effect, multi-band structure, magnetism, Fermi surface nesting and other uncontrollable interactions.
The physics in cold atom system is extremely clear in this sense.

The different driving mechanisms lead to completely different physical results. In solid materials\cite{Vitctor, Agterberg, Dim, Kaur, Matsunaga08},
the broadening of the FFLO phase in discussed in the temperature - magnetic field ($H-T$) plane.
In Ref.\onlinecite{Vitctor} the broadening of FFLO
phases mainly comes from the increase of H$_{2c}$ (the second critical field between superconductor and normal states) in solid state materials.
In contrast,  the broadening of FFLO phases in our work comes from the decrease of the critical Zeeman field between BCS superfluids and FFLO superfluids,
see Fig. \ref{fig-phase}d and Fig. 1d in Ref. \onlinecite{Zhen}.

\subsection{Different dimensionality}

In solid materials, the FFLO phase can only be observed in low dimensional systems, in which an external magnetic field parallel to the
sample surface can effectively suppress the orbital effect. The disadvantage is that the fluctuating effect is also significant in low dimensional
systems, which is probably one of the basic reasons that why FFLO phase is not observed in practical experiments. In cold atom systems, the orbital effect
is independent of the dimensionality of the system because because the Zeeman field is purely induced by Raman coupling and detuning of hyperfine states.
Therefore the FFLO phase can not only be observed in low dimensional systems, but also in three dimensional system\cite{Zhen}. The cold atom platform
has the additional advantage of disorder free.

\subsection{BCS limit versus BCS-BEC crossover physics}

In solid materials, $h\ll E_F$, and the relevant FFLO physics occurs in the BCS limit. In cold atom systems for the parameter
regime $h \sim E_F$, the FFLO phase can be observed in the strong coupling regime, which can be tuned by Feshbach resonance
($E_b$ in this work, and scattering length in Ref. \onlinecite{Zhen}). As a result, the relevant interesting physics in
cold atom system is the BEC-BCS crossover.

\subsection{Different realistic experimental conditions and concerns}

The solid state systems and ultra-cold atom systems have very different constraints for the
experimental realization and observation of FFLO phases. The following is a comparison:
Solid state systems: 1) Temperature is not an issue because of the large Fermi energy; 2)
Disorder is very important and its role is still not well understood (see Ref. \cite{Patrick} for the influence of
disorder on FFLO states); 3) The FFLO phase is hard to observe directly; 4) The scattering effect and associated
lifetimes are crucial for FFLO phase. Cold atomic systems: 1) Temperature is important because current experimentally reachable
temperature is around 0.05$E_F$. Therefore the energy difference between FFLO state
and BCS excited state is important, which is shown to be large in our scheme and is one major
advantage of our proposal; 2) Disorder free; 3) The FFLO phase can be observed directly in time
of flight images; 4) The system is very stable, and the lifetime issue is unimportant. The experimental tools in these materials
are also quite different.  In solids, the thermal conductivity \cite{Capan}, specific heat\cite{Bianchi}, nuclear magnetic resonance\cite{NMR1, NMR2, NMR3},
and ultrasound velocities\cite{Ultrasound} are generally used to study the anomalous properties of FFLO phase, while in cold atoms, the time-of-flight imaging can be
directly used to probe the pairing momentum and associated Fermi surface asymmetry. In this sense, the cold atom platform may provide the most convincing evidence 
for the formation of finite momentum pairing.

\section{Conclusion}

\label{sec-conclusion}

To summarize, in this paper we study the possible FFLO phase in SO coupled
degenerate Fermi gases with in-plane Zeeman fields. We show that the
parameter region for the FFLO phase can be greatly enlarged due to the
deformation of the Fermi surface. The emergence of the FFLO phase is
explained from different angles. The properties of the BCS superfluid, FFLO
phase and normal gas have also been discussed and their measurement through
the time-of-flight imaging is presented. Our results indicate that the
deformation of the Fermi surface provides a more efficient method to
generate the FFLO phase. Because the SO coupling has been realized in Bose
\cite{Spielmana, Spielmanb, Pan,Qu} and Fermi \cite{Jing,Zwierlein12} cold
atom gases in experiments, where the in-plane Zeeman field can be naturally
created \cite{Spielmana, Spielmanb, Pan, Jing,Zwierlein12} and tuned, we
expect the idea in this work may provide a path for elucidating the
long-standing problem about FFLO phases in experiments in the near future.

\textit{Acknowledgement:} Z.Z., Y.Z., X.Z., and G.G. are supported by the
National 973 Fundamental Research Program (Grant No. 2011cba00200), the
National Natural Science Foundation of China (Grant No. 11074244 and No.
11274295).  C.Z. is supported by ARO (W911NF-12-1-0334), DARPA-YFA
(N66001-10-1-4025), and NSF-PHY (1104546).  M.G is supported in part by Hong
Kong RGC/GRF Project 401512, the Hong Kong Scholars Program (Grant No. XJ2011027)
and the Hong Kong GRF Project 401113.


\begin{thebibliography}{99}
\bibitem{BCS} J. Bardeen, L. N. Cooper, and J. R. Schrieffer, Phys. Rev.
\textbf{106}, 162 (1957).

\bibitem{Ferrell64} P. Fulde, and R. A. Ferrell, Phys. Rev. \textbf{135},
A550 (1964).

\bibitem{Larkin64} A. I. Larkin, and Yu. N. Ovchinnikov, Zh. Eksp. Teor.
Fiz. \textbf{47}, 1136 (1964).

\bibitem{Larkin65} A. I. Larkin, and Yu. N. Ovchinnikov, Sov. Phys. JETP
\textbf{20}, 762 (1965).

\bibitem{Buzdin} A. Buzdin, Y. Matsuda, and T. Shibauchi, Europhys. Lett.,
\textbf{80}, 67004 (2007).

\bibitem{Croitoru} M. D. Croitoru, M. Houzet and A. I. Buzdin, Phys. Rev.
Lett. \textbf{108}, 207005 (2012).

\bibitem{Yuji} Y. Matsuda, and H. Shimahara, J. Phys. Soc. Jpn. \textbf{76},
051005 (2007).

\bibitem{Gloos} K. Gloos, R. Modler, H. Schimanski, C. D. Bredl, C. Geibel,
F. Steglich, A. I. Buzdin, N. Sato, and T. Komatsubara, Phys. Rev. Lett.
\textbf{70}, 501 (1993).

\bibitem{Bianchi} A. Bianchi, R. Movshovich, C. Capan, P. G. Pagliuso, and
J. L. Sarrao, Phys. Rev. Lett. \textbf{91}, 187004 (2003).

\bibitem{NMR1} V. F. Mitrovic, M. Horvatic, C. Berthier, G. Knebel, G. Lapertot, and J. Flouquet, Phys. Rev. Lett. \textbf{97}, 117002 (2006).
\bibitem{NMR2} B.-L. Young, R. R. Urbano, N. J. Curro, J. D. Thompson, J. L. Sarrao, A. B. Vorontsov, and M. J. Graf, Phys. Rev. Lett. \textbf{98}, 036402 (2007).
\bibitem{NMR3} K. Kakuyanagi, M. Saitoh, K. Kumagai, S. Takashima, M. Nohara, H. Takagi, and Y. Matsuda, Phys. Rev. Lett. \textbf{94}, 047602 (2005).

\bibitem{Ultrasound} T. Watanabe, Y. Kasahara, K. Izawa, T. Sakakibara, and Y. Matsuda, T. Hanaguri, H. Shishido, R. Settai, and Y. Onuki, Phys. Rev. B \textbf{70}, 020506(R) (2004).

\bibitem{Singleton} J. Singleton, J. A. Symington, M-S Nam, A. Ardavan, M.
Kurmoo, and P. Day, J. Phys.: Condens. Matter \textbf{12}, L641 (2000).

\bibitem{Lortz} R. Lortz, Y. Wang, A. Demuer, P. H. M. B\"ottger, B. Bergk,
G. Zwicknagl, Y. Nakazawa, and J. Wosnitza, Phys. Rev. Lett. \textbf{99},
187002 (2007).

\bibitem{Casalbuoni} R. Casalbuoni and G. Narduli, Rev. Mod. Phys. \textbf{76%
}, 263 (2004).

\bibitem{Alford} M. G. Alford, K. Rajagopal, T. Schaefer, A. Schmitt, Rev.
Mod. Phys. \textbf{80}, 1455 (2008).

\bibitem{Capan} C. Capan, A. Bianchi, R. Movshovich, A. D. Christianson,
A. Malinowski, M. F. Hundley, A. Lacerda, P. G. Pagliuso, and J. L. Sarrao, Phys. Rev. B \textbf{70}, 134513 (2004).

\bibitem{Houzet} M. Houzet and A. Buzdin, Phys. Rev. B \textbf{63}, 184521
(2001).

\bibitem{Alsam} L. G. Alsamzov, Zh. Eksp. Teor. Fiz. \textbf{55}, 1477
(1968) [Sov. Phys. JETP \textbf{28}, 773 (1969)].

\bibitem{JXZhu} A. V. Balatsky, I. Vekhter, J.-X. Zhu, Rev. Mod. Phys.
\textbf{78}, 373 (2006).

\bibitem{Feshbach1} M. W. Zwierlein, C. A. Stan, C. H. Schunck, S. M. F.
Raupach, A. J. Kerman, and W. Ketterle, Phys. Rev. Lett. \textbf{92}, 120403
(2004).

\bibitem{Feshbach2} C. Chin, R. Grimm, P. Julienne, and E. Tiesinga, Rev.
Mod. Phys. \textbf{82}, 1225 (2010).

\bibitem{Feshbach3} S. Giorgini, L. P. Pitaevskii, and S. Stringari, Rev.
Mod. Phys. \textbf{80}, 1215 (2008).

\bibitem{disorder1} P. Horak, J. Y. Courtois, G. Grynberg, Phys. Rev. A
\textbf{58}, 3953 (1998).

\bibitem{disorder2} L. Guidoni, C. Trich\'e, P. Verkerk, and G. Grynberg,
Phys. Rev. Lett. \textbf{79}, 3363 (1997).

\bibitem{disorder3} L. Sanchez-Palencia and M. Lewenstein, Nature Physics
\textbf{6}, 87 (2010).

\bibitem{Zwierlein05} M.W. Zwierlein, J. R. Abo-Shaeer, A. Schirotzek, C.H.
Schunck, and W. Ketterle, Nature \textbf{435}, 1047 (2005).

\bibitem{Sheehy} D. E. Sheehy and L. Radzihovsky, Phys. Rev. Lett. \textbf{96%
}, 060401 (2006).

\bibitem{TOF2} Y. L. Loh and N. Trivedi, Phys. Rev. Lett. \textbf{104},
165302 (2010).

\bibitem{Hu} H. Hu and X.-J. Liu, Phys. Rev. A \textbf{73}, 051603(R) (2006).

\bibitem{Zwierlein06a} M. W. Zwierlein, A. Schirotzek, C. H. Schunck, and W.
Ketterle, Science \textbf{311}, 492 (2006).

\bibitem{Zwierlein06b} M. W. Zwierlein, C. H. Schunck, A. Schirotzek, and W.
Ketterle, Nature \textbf{442}, 54 (2006).

\bibitem{Hulet1} G. B. Partridge, W. Li, R. I. Kamar, Y.-A. Liao and R. G.
Hulet, Science, \textbf{311}, 503 (2006).

\bibitem{Vicent} W. V. Liu, F. Wilczek, Phys. Rev. Lett. \textbf{90}, 047002
(2003).

\bibitem{SHE1} J. Wunderlich, B. Kaestner, J. Sinova, and T. Jungwirth, Phys. Rev. Lett. \textbf{94}, 047204 (2005).

\bibitem{SHE2}  C. L. Kane and E. J. Mele , Phys. Rev. Lett. \textbf{95}, 226801 (2005).

\bibitem{AHE1} T. Jungwirth, Qian Niu, and A. H. MacDonald, Phys. Rev. Lett. \textbf{88}, 207208 (2002).

\bibitem{AHE2}  Naoto Nagaosa, Jairo Sinova, Shigeki Onoda, A. H. MacDonald, N. P. Ong, Rev. Mod. Phys. \textbf{82}, 1539 (2010).

\bibitem{TI1} Haijun Zhang, Chao-Xing Liu, Xiao-Liang Qi, Xi Dai, Zhong Fang and Shou-Cheng Zhang, Nat. Phys. \textbf{5}, 438 (2009).

\bibitem{TI2} Xiao-Liang Qi and Shou-Cheng Zhang, Rev. Mod. Phys. \textbf{83}, 1057 (2011).

\bibitem{TI3} M. Z. Hasan and C. L. Kane, Rev. Mod. Phys. \textbf{82}, 3045 (2010).

\bibitem{Winkle} R. Winkler, Spin-Orbit Coupling Effects in Two-Dimensional
Electron and Hole Systems, Springer Tracts in Modern Physics Vol. \textbf{191}
(Springer-Verlag, Berlin, 2003).

\bibitem{Ruseckas} J. Ruseckas, G. Juzeliuna1, P. \"{o}hberg, and M.
Fleischhauer, Phys. Rev. Lett. \textbf{95}, 010404 (2005).

\bibitem{Xiongjun} X.-J. Liu, M. F. Borunda, X. Liu, and J. Sinova, Phys.
Rev. Lett. \textbf{102}, 046402 (2009).

\bibitem{xiongjun2} X.-J. Liu, X. Liu, L. C. Kwek, and C. H. Oh, Phys. Rev.
Lett. \textbf{98}, 026602 (2007).

\bibitem{CWZhang} C. Zhang, Phys. Rev. A \textbf{82}, 021607 (2010).

\bibitem{Campbell} D. L. Campbell, G. Juzeliunas, and I. B. Spielman, Phys.
Rev. A \textbf{84}, 025602 (2011).

\bibitem{Jay} J. D. Sau, R. Sensarma, S. Powell, I. B. Spielman, and S. Das
Sarma, Phys. Rev. B \textbf{83}, 140510(R) (2011).

\bibitem{non-Abelian} J. Dalibard, F. Gerbier, G. Juzeliunas, and P. \"{O}%
hberg, Rev. Mod. Phys. \textbf{83}, 1523 (2011).

\bibitem{Spielmana} Y.-J. Lin, K. Jimenez-Garcia, and I. B. Spielman, Nature
\textbf{471}, 83 (2011).

\bibitem{Spielmanb} Y.-J. Lin, R. L. Compton, A. R. Perry, W. D. Phillips,
J. V. Porto, and I. B. Spielman, Phys. Rev. Lett. \textbf{102}, 130401 (2009).

\bibitem{Pan} J.-Y. Zhang \textit{et al.}, Phys. Rev. Lett. \textbf{109}, 115301
(2012).

\bibitem{Qu} Chunlei Qu, Chris Hamner, Ming Gong, Chuanwei Zhang, and Peter Engels, Phys. Rev. A \textbf{88}, 021604 (2013).

\bibitem{Jing} P. Wang, Z.-Q. Yu, Z. Fu, J. Miao, L. Huang, S. Chai, H.
Zhai, J. Zhang, Phys. Rev. Lett. \textbf{109}, 095301 (2012).

\bibitem{Zwierlein12} L. W. Cheuk, A. T. Sommer, Z. Hadzibabic, T. Yefsah,
W. S. Bakr, M. W. Zwierlein, Phys. Rev. Lett. \textbf{109}, 095302 (2012).

\bibitem{Zhen} Z. Zheng, M. Gong, X. Zou, C. Zhang and G.-C. Guo, Phys. Rev. A \textbf{87}, 031602(R) (2013)

\bibitem{SOFF1} Xiang-Fa Zhou, Guang-Can Guo, Wei Zhang, Wei Yi, Phys. Rev. A \textbf{87}, 063606 (2013).
\bibitem{SOFF2} Fan Wu, Guang-Can Guo, Wei Zhang, Wei Yi, Phys. Rev. Lett. \textbf{110}, 110401 (2013).
\bibitem{SOFF3} Lin Dong, Lei Jiang, Hui Hu, Han Pu, Phys. Rev. A \textbf{87}, 043616 (2013).
\bibitem{SOFF4} Xia-Ji Liu, Hui Hu, Phys. Rev. A \textbf{87}, 051608(R) (2013).
\bibitem{SOFF5} M. Iskin, Phys. Rev. A \textbf{88}, 013631 (2013).
\bibitem{SOFF6} Hui Hu, and Xia-Ji Liu, New J. Phys. \textbf{15}, 093037 (2013).

\bibitem{Martiyanov} K. Martiyanov, V. Makhalov, and A. Turlapov, Phys. Rev. Lett. \textbf{105}, 030404 (2010).

\bibitem{Gang} G. Chen, M. Gong, and C. Zhang, Phys. Rev. A \textbf{85}, 013601 (2012).

\bibitem{Schnyder} A. P. Schyder, S. Ryu, A. Furusaki, and A. W. W. Ludwig, Phys. Rev. B \textbf{78}, 195125 (2008).

\bibitem{NC} Chunlei Qu, Zhen Zheng, Ming Gong, Yong Xu, Li Mao, Xubo Zou, Guangcan Guo, Chuanwei Zhang, Nature Communications \textbf{4}, 3710 (2013).

\bibitem{He06} L. He, M. Jin and P. Zhuang, Phys. Rev. B \textbf{74}, 024516 (2006).

\bibitem{Conduit} G. J. Conduit, P. H. Conlon, B. D. Simons, Phys. Rev. A
\textbf{77}, 053617 (2008).

\bibitem{CWZhang2} C. Zhang, S. Tewari, R. M. Lutchyn, and S. Das Sarma, Phys. Rev. Lett. \textbf{101}, 160401 (2008).
\bibitem{Gong11} M. Gong, S. Tewari and C. Zhang, Phys. Rev. Lett. \textbf{107}, 195303 (2011).

\bibitem{Gong12} M. Gong, G. Chen, S. Jia, and C. Zhang, Phys. Rev. Lett. \textbf{109}, 105302 (2012).

\bibitem{Iskin2} M. Iskin, Phys. Rev. A. \textbf{85}, 013622 (2012).

\bibitem{Altman} E. Altman, E. Demler, and M. D. Lukin, Phys. Rev. A \textbf{70},
013603 (2004).

\bibitem{Koponen} T. K. Koponen, T. Paananen, J-P Martikainen, and P. T\"{o}%
rma, Phys. Rev. Lett. \textbf{99}, 120403 (2007).

\bibitem{Koponen2} T. K. Koponen, T. Paananen, J-P Martikainen, M. r.
Bakhtiari, and P. T\"{o}rma, New. J. Phys. \textbf{10}, 045014 (2008).

\bibitem{Stein} J. Steinhauer, N. Katz, R. Ozeri, N. Davidson, C. Tozzo, F. Dalfovo, Phys. Rev. Lett. \textbf{90}, 060404 (2003).

\bibitem{Duan} L.-M. Duan, Phys. Rev. Lett. \textbf{96}, 103201 (2006).

\bibitem{Vitctor} V. Barzykin and L. Gor'kov, Phys. Rev. Lett. \textbf{89}, 227002 (2002).

\bibitem{Yip02} S. K. Yip , Phys. Rev. B \textbf{65}, 144508 (2002).

\bibitem{Agterberg} D. F. Agterberg, Phyica C \textbf{387}, 13 (2003).

\bibitem{Dim} Ol'ga Dimitrova and M. V. Feigel'man, Phys. Rev. B \textbf{76}, 014522 (2007).

\bibitem{Kaur} D. F. Agterberg and R. P. Kaur, Phys. Rev. B \textbf{75}, 064511 (2007).

\bibitem{Matsunaga08} Yuichi Matsunaga, Norihito Hiasa, and Ryusuke Ikeda, Phys. Rev B \textbf{78}, 220508(R) (2008).

\bibitem{Oka06} Masatosi Oka, Masanori Ichioka, and Kazushige Machida, Phys. Rev. B \textbf{73}, 214509 (2006).

\bibitem{Patrick} Karen Michaeli, Andrew C. Potter, and Patrick A. Lee, Phys. Rev. Lett. \textbf{108}, 117003 (2012).

\bibitem{Yuan} H. Q. Yuan, D. F. Agterberg, N. Hayashi, P. Badica, D. Vandervelde, K. Togano, M. Sigrist, and M. B. Salamon, Phys. Rev. Lett. \textbf{97}, 017006 (2006).
\bibitem{Yokoyama} Takehito Yokoyama, Seiichiro Onari, Yukio Tanaka, J. Phys. Soc. Jpn. \textbf{77}, 064711 (2008).
\bibitem{Lee} K.-W. Lee and W. E. Pickett, Phys. Rev. B \textbf{72}, 174505 (2005).
\bibitem{Tada}Y. Tada, N. Kawakami, and S. Fujimoto, Phys. Rev. Lett. \textbf{101}, 267006 (2008).
\bibitem{Frigeri} P. Frigeri, D.F. Agterberg, A. Koga, M. Sigrist, Phys. Rev. Lett. \textbf{92}, 097001 (2004).

\end{thebibliography}
\end{document}